\documentclass[aps,prd,11pt,superscriptaddress,amssymb,amsmath,nofootinbib]{revtex4-1}

\usepackage{hhline}
\usepackage{color}
\usepackage{epsfig}
\usepackage{graphicx}
\usepackage{float}
\usepackage{epstopdf}
\usepackage{hyperref}
\usepackage[usenames,dvipsnames]{xcolor}
\usepackage{multirow,booktabs}
\usepackage[normalem]{ulem}
\usepackage[d]{esvect}
\usepackage[misc,geometry]{ifsym}
\usepackage{harpoon}

\begin{document}

	\title{Influences of non-standard interactions on PeV neutrino events with and without a $L_{\alpha}-L_{\beta}$ symmetry }
	\author{Qiu-Xia Yi}\email{1020221818@glut.edu.cn}
\author{Ya-Ru Wang}\email{1020231946@glut.edu.cn}
	\author{Shu-Jun Rong\Letter}\email{rongshj@glut.edu.cn}
	\affiliation{College of Physics and Electronic Information Engineering, Guilin University of Technology, Guilin, Guangxi 541004, China}

\begin{abstract}
The recently reported astrophysical neutrinos events in the TeV-PeV energy range open a winder to explore new physics at energy frontiers.
In this paper, we examine effects of non-standard interactions (NSIs) on the PeV neutrinos events. We consider NSIs with and without a gauge symmetry $L_{\alpha}$ - $L_{\beta}$.
We find that, for typical $\mu^{\pm}$ damping and $\pi^{\pm}$ decay sources, the NSI with an extra gauge symmetry has more noticeable effects on the PeV events. Therefore, the
detection of the events in the upcoming experiments could set stringent constraints on the NSI parameters in the  $L_{\alpha}$ - $L_{\beta}$ symmetric case.

\end{abstract}
	
\maketitle
	
\section{Introduction}
Searching for signals of physics beyond the standard model (SM) is an active field in particle physics, which is mainly focused on the precision or energy frontier.
Recent observed astrophysical neutrino events in the TeV-PeV energy range\cite{IceCube:2013low,IceCube:2013cdw,IceCube:2014stg,IceCube:2016umi,IceCube:2021uhz} provide a probe to explore new physics such as Lorentz invariance violation\cite{Colladay:1998fq,Coleman:1998ti,Kostelecky:2003cr,Roberts:2021vsi,Zhang:2022svg}, neutrino decay\cite{Beacom:2002vi,Meloni:2006gv,Baerwald:2012kc,Pagliaroli:2015rca,Bustamante:2016ciw,Denton:2018aml,Huang:2024tbo,Valera:2024buc}, pseudo-Dirac neutrinos\cite{Kobayashi:2000md,Beacom:2003eu,Carloni:2022cqz,Rink:2022nvw,Franklin:2023diy}, non-unitary leptonic mixing matrix\cite{Antusch:2014woa,Blennow:2023mqx,Aloni:2022ebm}, non-standard interaction (NSI)\cite{Huitu:2016bmb,MINOS:2016sbv,Esteban:2018ppq,Borexino:2019mhy,Masud:2021ves,IceCubeCollaboration:2021euf,Bakhti:2020fde,Brahma:2022xld,LazoPedrajas:2024qlf}. Based on a direct modification of the Lagrangian of SM, NSIs of neutrinos  could change their flavor oscillation. The  NSI parameters are constrained by atmospheric neutrinos
\cite{Fornengo:2001pm,IceCubeCollaboration:2021euf}, long-baseline oscillation\cite{MINOS:2016sbv,Bakhti:2020fde}, and  the coherent elastic neutrino-nucleus scattering experiments like COHERENT\cite{Liao:2024qoe}. In this paper, we investigate the constrains on NSI parameters with the updated data of IceCube\cite{IceCube:2023sov} and examine the effects of NSI on astrophysical neutrino events at PeV energies.
	
The influence of NSI on neutrino oscillation is parameterized by a $3\times3$ Hamiltonian matrix. The matrix contains 9 real parameters, which makes predictions on the deviation from the standard oscillation complex at high energies. In the case where the NSI satisfies some symmetry, the form of the Hamiltonian can be simplified. A well-known
example is the gauge symmetry $L_{\alpha}-L_{\beta}$\cite{He:1991qd}, where $\alpha, \beta$ represents $e, \mu, \tau$. Furthermore, if the mass of the extra gauge boson is tiny,
the propagation of neutrinos could be impacted by a long-range potential\cite{Bustamante:2018mzu,Coloma:2020gfv,Agarwalla:2023sng,Agarwalla:2024ylc}. In this case, the Hamiltonian matrix of NSI is diagonal in the flavor bases.
The long-range interaction parameters are constrained by the results of neutrino oscillations experiments and IceCube observations. For astrophysical neutrinos, the effect of the long-range potential on the flavor ratio at Earth is mainly considered. However, because of the limitation of the resolution of neutrino flavor at IceCube, the constraint on the flavor ratio is loose at present. In this paper, we examine effects of the NSI satisfying a $L_{\alpha}-L_{\beta}$ symmetry on the PeV astrophysical neutrinos to complement the work on the long-range interaction influences. To be specific,
we study the impacts of the NSI potential on the Glashow resonance events\cite{Glashow:1960zz,Barger:2014iua,Huang:2019hgs,Xu:2022svm} and $\nu_{\tau}$ events\cite{IceCube:2024nhk}. Although the number of the events is rare by now, several PeV events in the upcoming experiments (e.g., P-ONE)\cite{P-ONE:2020ljt}, IceCube-Gen2\cite{IceCube-Gen2:2020qha}) may give strong restrictions on the NSI potential. In addition, we
consider the effects of NSI without a $L_{\alpha}-L_{\beta}$ symmetry on the events as a comparison.

The paper is organised as follows. In Sec. \uppercase\expandafter{\romannumeral2}, the neutrino flavor transition probability on the bases of NSI is introduced.
In Sec. \uppercase\expandafter{\romannumeral3}, the impacts of NSI on PeV neutrinos, including the flavor ratio, the flux, and the events number, are examined. Finally, a conclusion is given.

\section{Flavor conversion probability}
The effective neutrino-matter interaction leading to a long-range potential is subject to three contributions, namely the Standard Model (SM) term mediated by the Z boson, the interaction of the gauge symmetry $L_{\alpha}$-$L_{\beta}$ mediated by a novel boson $Z^{'}_{\alpha\beta}$, and the contribution of mixing between the Z and $Z^{'}_{\alpha\beta}$\cite{Agarwalla:2023sng}. It reads as follow,
\begin{equation}\label{eq:1}
	L_{eff}=L_{SM}+L_{Z^{'}}+L_{mix}.
\end{equation}
The first item on the right-hand is the contribution from the SM
\begin{equation}\label{eq:2}
	L_{SM}=\frac{e}{sin\theta_{W}cos\theta_{W}}Z_{\mu}[-\frac{1}{2}\overline{l}_{\alpha}\gamma^{\mu}P_{L}l_{\alpha}+\frac{1}{2}\overline{\nu}_{\alpha}\gamma^{\mu}P_{L}\nu_{\alpha}+\frac{1}{2}\overline{u}\gamma^{\mu}P_{L}u-\frac{1}{2}\overline{d}\gamma^{\mu}P_{L}d~].
\end{equation}
The second term describes neutrino-matter interactions through the new mediator $Z^{'}_{\alpha\beta}$\cite{Agarwalla:2023sng}, i.e.,
\begin{equation}\label{eq:3} L_{Z^{'}}=g^{'}_{\alpha\beta}Z^{'}_{\sigma}(\overline{l}_{\alpha}\gamma^{\sigma}l_{\alpha}-\overline{l}_{\beta}\gamma^{\sigma}l_{\beta}+\overline{\nu}_{\alpha}\gamma^{\sigma}P_{L}\nu_{\alpha}-\overline{\nu}_{\beta}\gamma^{\sigma}P_{L}\nu_{\beta}).
\end{equation}
It shows the contribution from the $L_{e}-L_{\mu}$ and $L_{e}-L_{\tau}$ gauge symmetries, and the corresponding interaction is noticeable due to a source rich of electrons.
 The third term generated from the mixing of the bosons $Z$ and $Z^{'}$ is written as\cite{Agarwalla:2023sng}
 \begin{equation}\label{eq:4}
 L_{mix}=-g^{'}_{\alpha\beta}(\xi-sin\theta_{W}\chi)\frac{e}{sin\theta_{W}cos\theta_{W}}J^{'}_{\rho}J^{\rho}_{3}.
\end{equation}
It only affects the results under the $L_{\mu}-L_{\tau}$ symmetry due to a neutron source\cite{Agarwalla:2023sng}.
On the bases  $L_{eff}$ and the assumption that the matter is isoscalar and electrically neutral, the total potential of the Earth, Moon, Sun, Milky Way and Cosmological should be considered\cite{Bustamante:2018mzu}, namely
\begin{equation}\label{eq:5}
	V_{\alpha\beta}=V_{\alpha\beta}^{E}+V_{\alpha\beta}^{M}+V_{\alpha\beta}^{S}+V_{\alpha\beta}^{MW}+V_{\alpha\beta}^{<COS>}.
\end{equation}
Since the number density of electrons and neutrons in the universe changes as the universe expands, the redshift-averaged potential is employed.
In this paper, we take a simplified view-point, and treat the total $V_{\alpha\beta}$ as a new-physics parameter to constrain by observations, irrespective of its specific origins.

Accordingly, the Hamiltonian matrix of neutrinos during their propagation is generalised as
\begin{equation}\label{eq:6}
	\mathbf{H}=\mathbf{H_{0}}+\mathbf{V_{\alpha\beta}}.
\end{equation}
The first term on the right-hand determines oscillations in vacuum and the standard neutrino-matter interactions, i.e., \cite{Wolfenstein:1977ue,Mikheyev:1985zog}
\begin{equation}\label{eq:7}
     \mathbf{H_{0}}=U
	\left(
	\begin{array}{ccc}
		0&0&0\\
		0&\frac{\Delta m^{2}_{21}}{2E}&0\\
		0&0&\frac{\Delta m^{2}_{31}}{2E}\\
	\end{array}
	\right)U^{\dagger}+
	\left(
	\begin{array}{ccc}
		V_{CC}&0&0\\
		0&0&0\\
		0&0&0\\
	\end{array}
	\right),
\end{equation}
where $U$ is the leptonic mixing matrix in vacuum, $V_{CC}=\sqrt{2}G_{F}N_{e}$ is the standard matter potential.
The second term working as the long-range potential is due to tiny masses of mediators of the new neutrino-matter interactions.

According to the analysis of the reference\cite{Agarwalla:2023sng}, the potentials of the form $L_{e}-L_{\mu}$ and $L_{e}-L_{\tau}$ bring similar results of flavor conversion probability. So the forms of $L_{e}-L_{\tau}$ and $L_{\mu}-L_{\tau}$ are considered in the following sections. As $\alpha, \beta =  e, \tau;  \mu, \tau$, the expressions of the term $\mathbf{V_{\alpha\beta}}$ are respectively
\begin{equation}\label{eq:8}
	\mathbf{V_{e\tau}}=
	\left(
	\begin{array}{ccc}
	V_{e\tau}~&~0~&~0~\\
	0&~0~&~0~\\
	0~&~0~&~-V_{e\tau}
	\end{array}
	\right);~~
		\mathbf{V_{\mu\tau}}=
	\left(
	\begin{array}{ccc}
	 0~&~0~&~0~\\
	 0~&~V_{\mu\tau}~&~0~\\
	 0~&~0~&~-V_{\mu\tau}
	\end{array}
	\right).
\end{equation}
Considering antineutrinos, $V^{ant}_{\alpha\beta}$=$-V_{\alpha\beta}$.

The $\mathbf{V_{\alpha\beta}}$ shown here is a special case derived from a NSI between neutrinos and matter.
In order to compare the effects of NSIs with and without a $L_{\alpha}-L_{\beta}$ symmetry, we introduce the non-diagonal potential from general NSIs, namely
\begin{equation}\label{eq:9}
	\mathbf{V_{e\tau}}^{(non)}=
	\left(
	\begin{array}{ccc}
		0~&~0~&~V_{e\tau}~\\
		0~&~0~&~0~\\
		V_{e\tau}~&~0~&~0
	\end{array}
	\right);~~
	\mathbf{V_{\mu\tau}}^{(non)}=
	\left(
	\begin{array}{ccc}
		0~&~0~&~0~\\
		0~&~0~&~ V_{\mu\tau}~\\
		0~&~V_{\mu\tau}~&~0
	\end{array}
	\right).
\end{equation}
For simplicity, $V_{\alpha\beta}$ still takes real values in this case.

Considering the limited energy resolution of neutrino telescopes which are only sensitive to the average probability, the $\nu_{\alpha} \rightarrow \nu_{\beta}$ transition probability reads
\begin{equation}\label{eq:10}
	\overline{P}^{N}_{\alpha\beta}=|U^{N}_{\alpha1}|^{2}|U^{N}_{\beta1}|^{2}+|U^{N}_{\alpha2}|^{2}|U^{N}_{\beta2}|^{2}+|U^{N}_{\alpha3}|^{2}|U^{N}_{\beta3}|^{2},
\end{equation}
where $U^{N}$ is derived from the diagonalization of the total Hamiltonian.

Note that the influence of $V_{CC}$ can be neglected in the case that the range of the new interaction is larger than the size of the Earth. Accordingly,
we focus on the case that $V_{\alpha\beta}$ is dominant in the $\mathbf{H}$ for the PeV neutrinos.

\section{ Constraints of potential parameters}
Now we examine the constraints on the NSI parameters.
As for the data sample, 4318 days of high-energy start-up events are used for the analysis, with a total of 164 updated events\cite{IceCube:2023sov}. Considering that there may be background events in the sample, namely atmospheric $\mu$, atmospheric neutrinos, and transient neutrinos\cite{IceCube:2016umi,Halzen:2016pwl,Halzen:2016thi}, we need a clean set of events to obtain more reliable analysis results. Based on the reason, the maximum likelihood analysis considers the shower events mainly generated by the  interaction of $\nu_{e}(\overline{\nu}_{e})$ and  $\nu_{\tau}(\overline{\nu}_{\tau})$\cite{IceCube:2015rro,IceCube:2015qii,Kopper:2017zzm,Denton:2018aml}, and sets the energy range of neutrinos to 60TeV-3PeV with a total of 64 data samples, see Fig.\ref{fig:1}.
\begin{figure}\label{fig:1}
	\centering
	\includegraphics[width=0.6\textwidth]{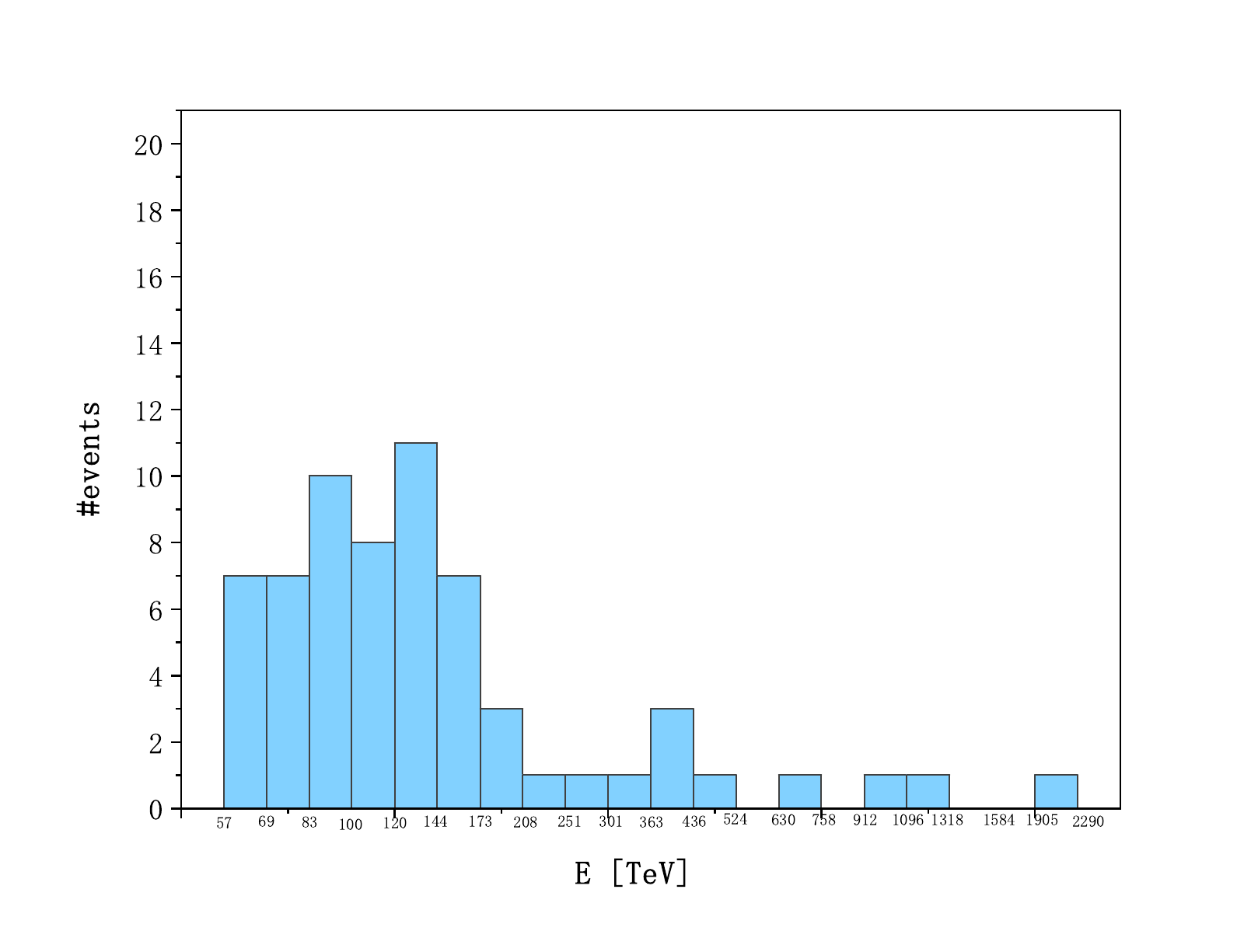}
	\caption{\label{fig:1}  The shower events number in the energy-range 60TeV-3PeV.}
	\label{fig:1}
\end{figure}

We employ a binned maximum likelihood method to constrain the parameters of the NSI potential in the neutrino propagation path. The expected number of events at each energy bin k is
\begin{equation}\label{eq:11}
	N_{k}(V_{\alpha\beta})=4\pi T(\int_{k}\Phi_{\nu_{e}+\overline{\nu}_{e}}(E,V_{\alpha\beta})A_{e,k}(E)dE+\int_{k}\Phi_{\nu_{\tau}+\overline{\nu}_{\tau}}(E,V_{\alpha\beta})A_{\tau,k}(E)dE),
\end{equation}
where T=4318 days is the time of data-collecting, $\int_{k}$ represents the integral over each interval k. $A_{\alpha,~k}(E)$ represents the effective area of the IceCube Collaboration corresponding to the $\alpha$-flavor neutrino at the k energy interval\cite{IceCube:2013low}, and $\Phi_{\nu_{\alpha}+\overline{\nu}_{\alpha}}(E,V_{\alpha\beta})=\Phi_{\nu+\overline{\nu}}(E)(\sum_{\beta}\overline{P}^{N}_{\alpha\beta}(E,V_{\alpha\beta})\phi^{s}_{\beta})$ with $\phi^{s}_{\beta}$ being the flavor ratio at the source of astrophysical neutrinos.
$\Phi_{\nu+\overline{\nu}}(E)$ indicates the entire diffuse flow, and its specific expression is as follow
\begin{equation}\label{eq:12}
	\Phi_{\nu+\overline{\nu}}(E)=\phi\times(\frac{E}{100~\rm TeV})^{-\gamma}\cdot10^{-18}~\rm GeV^{-1} cm^{-2} s^{-1} sr^{-1}.
\end{equation}
The best fit values and 68\% confidence level of energy-spectrum parameters in the all-sky model are listed in Table~\uppercase\expandafter{\romannumeral1}\cite{IceCube:2020wum}.
\begin{table}
	\caption{\label{Tab:1} The best fit values and 68\% confidence level of energy-spectrum parameters in the all-sky model\cite{IceCube:2020wum}.}
	\centering
	\begin{tabular}{ccc}
		\hline
		~~~~~~parameters~~~~~~&~~~~~~best fit value~~~~~~&~~~~~~68\% confidence level(C.L.)~~~~~~\\
		\hline
		~~~~~~$\phi$~~~~~~&~~~~~~6.37~~~~~~&~~~~~~~~4.75~-~7.83~~~~~~\\
		~~~~~~$\gamma$~~~~~~&~~~~~~2.87~~~~~~&~~~~~~~~2.68~-~3.08~~~~~~\\
		\hline
	\end{tabular}
\end{table}
The number of shower events in the energy interval k is denoted by $\overline{N_{k}}$. It is further assumed that the expected number of shower events $N_{k}({V_{\alpha\beta}})$ in different energy intervals k obey the poisson distribution, namely
\begin{equation}\label{eq:13}
	P[\overline{N_{k}}|N_{k}(V_{\alpha\beta})]=\frac{N_{k}(V_{\alpha\beta})^{\overline{N_{k}}}}{\overline{N_{k}}!}\exp\{-N_{k}(V_{\alpha\beta})\}.
\end{equation}
Each likelihood function is
\begin{equation}\label{eq:14}
	L(V_{\alpha\beta})=\prod_{k}P[~\overline{N_{k}}|N_{k}(V_{\alpha\beta})].
\end{equation}
Taking the logarithm of both sides of the above equation, we can obtain the following expression
\begin{equation}\label{eq:15}
	\ln[L(V_{\alpha\beta})]=\sum_{k}\ln\{~P[~\overline{N_{k}}|N_{k}(V_{\alpha\beta})]\}.
\end{equation}

The value of the function $\ln[L(V_{\alpha\beta})]$ is dependent on the leptonic mixing parameters and the energy-spectrum parameters of the diffuse flux.
However, we find that the mixing parameters and the spectrum index $\gamma$ moderately impact the magnitude of $\ln[L(V_{\alpha\beta})]$,  and the normalization parameter $\phi$
determines the characteristic of dependence of $\ln[L(V_{\alpha\beta})]$ on $V_{\alpha\beta}$. Therefore, the best fit data of the mixing parameters with  the normal mass-ordering (NO) from the global fit  analysis NuFIT5.0 published by PDG\cite{Esteban:2020cvm} and the best fit value of the spectrum index in Tab.\ref{Tab:1} are taken here. The influence of normalization parameter $\phi$ on $\ln[L(V_{\alpha\beta})]$ is averaged with a Monte Carlo integration method. Based on the set-up of the nuisance parameters, the behaviors of $-\ln[L(V_{\alpha\beta})]$ of the models are shown in Fig.\ref{fig:2}-\ref{fig:3}.  Accordingly, the upper limits (95\% C.L.) on the potentials are obtained, see Tab.\ref{Tab:2}-\ref{Tab:3}. We note that although the statistic procedure employed here is simple, the updated sample data of IceCube can set more stringent constraints on the
potential parameters in comparison with the results from the long-baseline experiments\cite{Agarwalla:2024ylc}.

\begin{figure}\label{fig:2}
	\centering
	\includegraphics[width=0.45\textwidth]{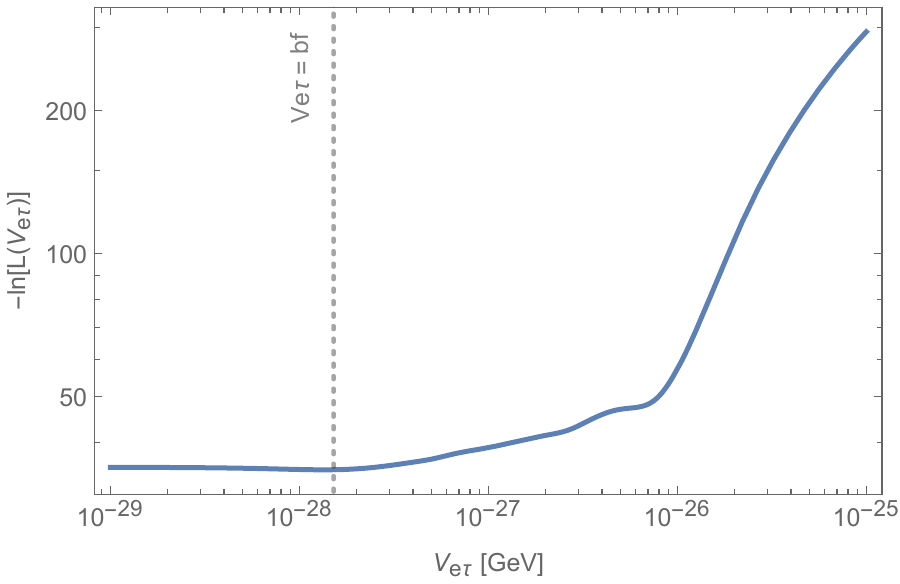}
	\hfill
	\includegraphics[width=0.45\textwidth]{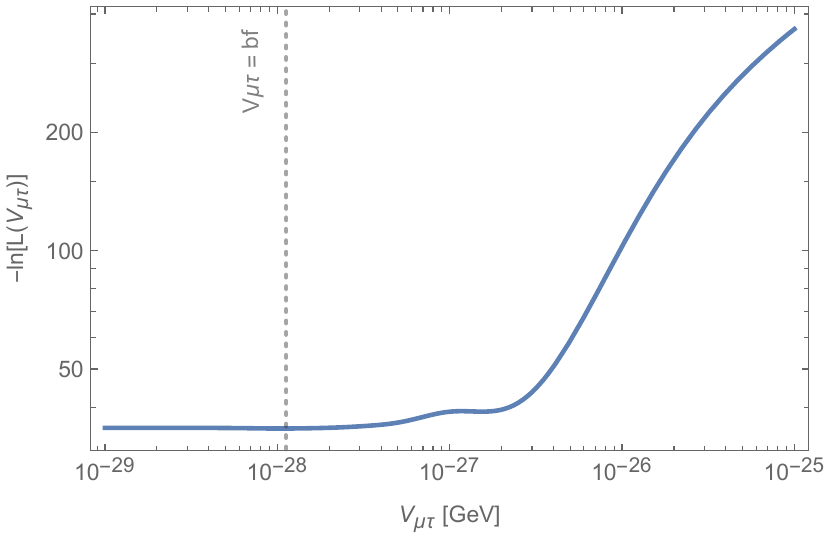}
	\hfill
	\includegraphics[width=0.45\textwidth]{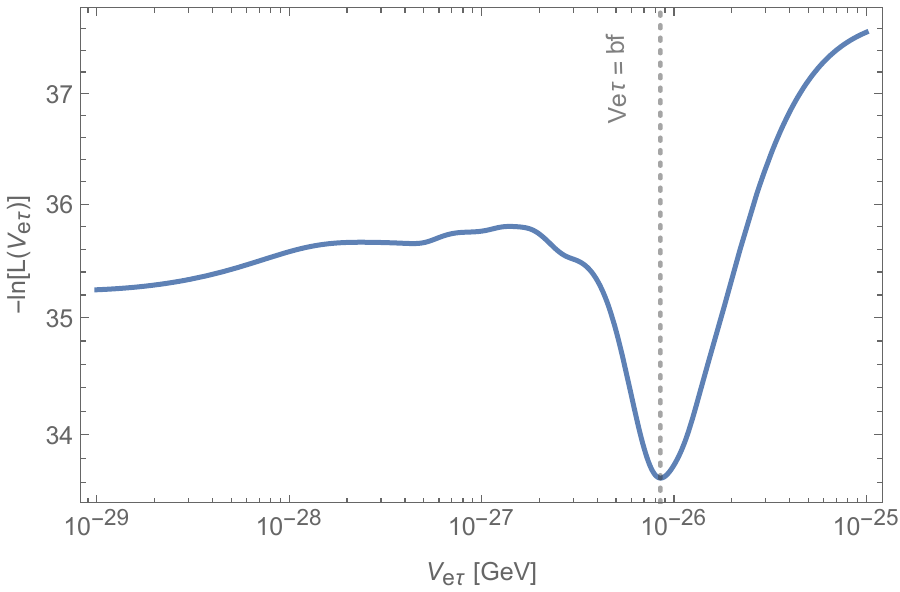}
	\hfill
	\includegraphics[width=0.45\textwidth]{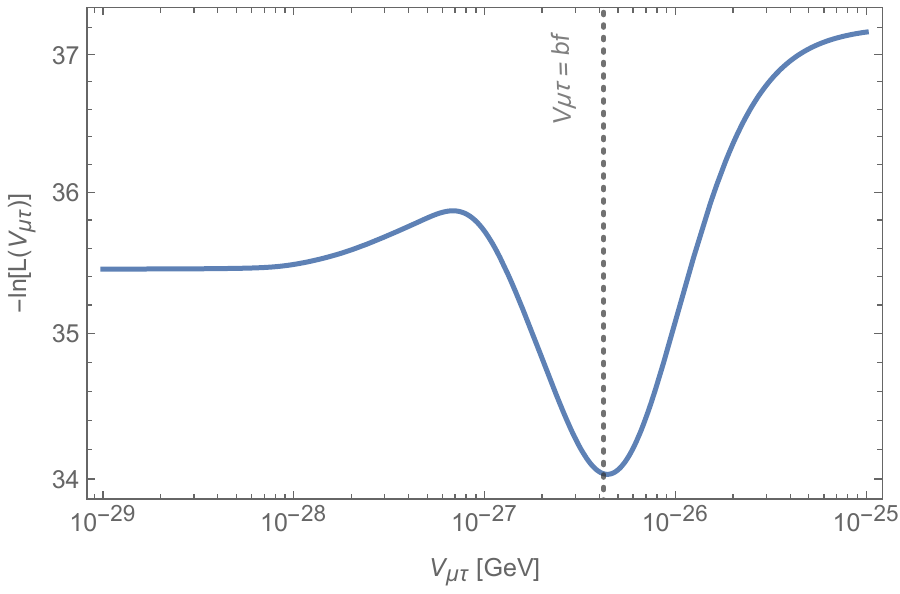}
	\hfill
	\caption{\label{fig:2} The dependence of the function $-\ln[~L(V_{\alpha\beta})~]$ on the parameter $V_{\alpha\beta}$ in the $L_{\alpha}-L_{\beta}$ symmetric model. The first row, left panel: $L_{e}-L_{\tau}$  model with $\mu^{\pm}$ damping source. The first row, right panel: $L_{\mu}-L_{\tau}$ model with $\mu^{\pm}$ damping source. The second row, left panel:  $L_{e}-L_{\tau}$ with $\pi^{\pm}$ decay source. The second row, right panel: $L_{\mu}-L_{\tau}$ with $\pi^{\pm}$ decay source. The flavor ratio for $\mu^{\pm}$ damping is $\phi^{S}_{\alpha}=(0,1,0)$, and ratio for $\pi^{\pm}$ decay is $\phi^{S}_{\alpha}=(1/3,2/3,0)$. }
	\label{fig:2}
\end{figure}
\begin{figure}\label{fig:3}
	\centering
	\includegraphics[width=0.45\textwidth]{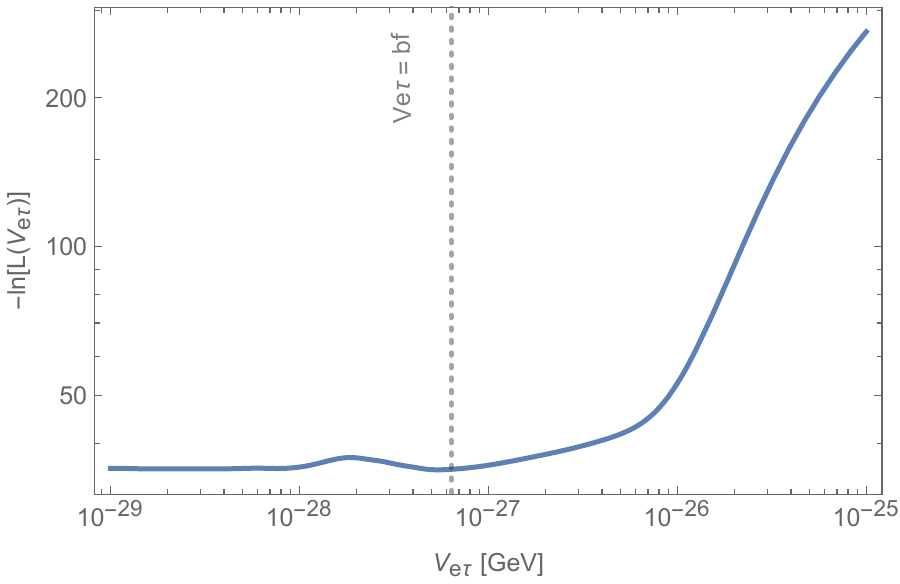}
	\hfill
	\includegraphics[width=0.45\textwidth]{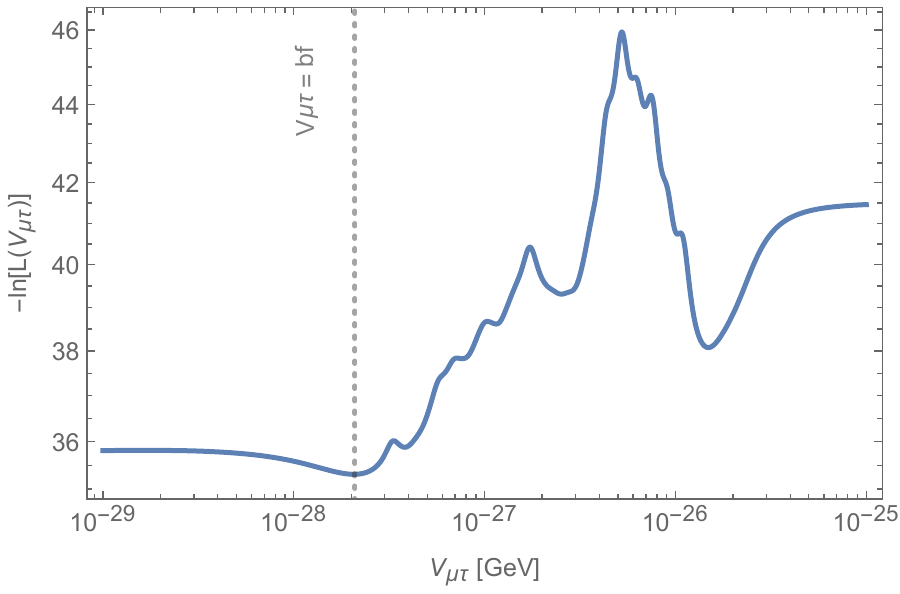}
	\hfill
	\includegraphics[width=0.45\textwidth]{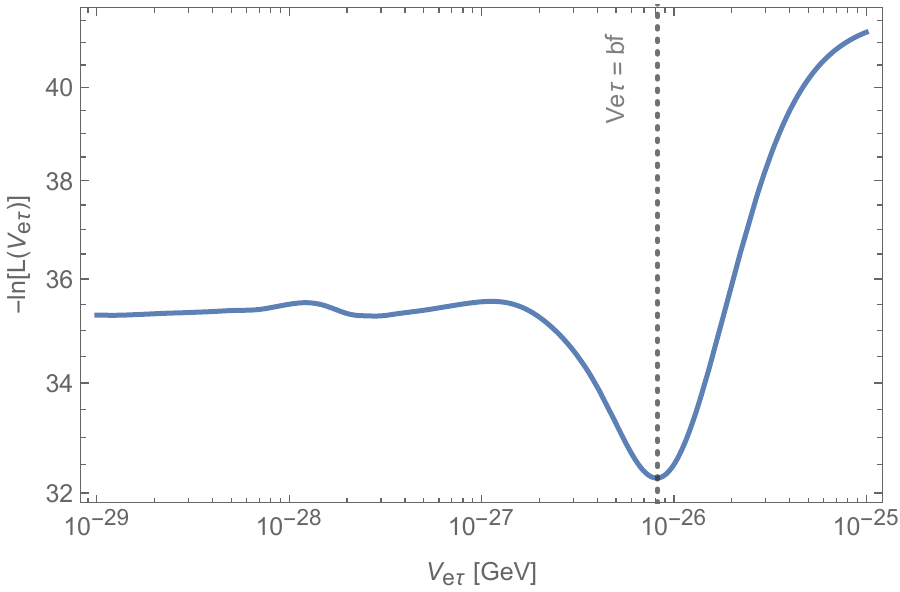}
	\hfill
	\includegraphics[width=0.45\textwidth]{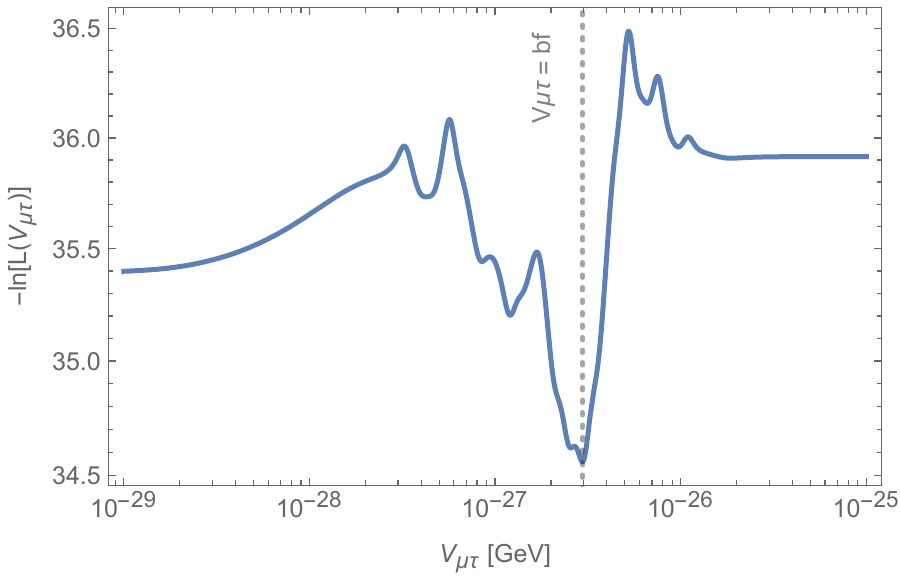}
	\hfill
	\caption{\label{fig:3} The dependence of the function $-\ln[~L(V_{\alpha\beta})~]$ on the parameter $V_{\alpha\beta}$ in the $L_{\alpha}-L_{\beta}$ breaking model. The first row, left panel: $L_{e}-L_{\tau}$  model with $\mu^{\pm}$ damping source. The first row, right panel: $L_{\mu}-L_{\tau}$  model with $\mu^{\pm}$ damping source. The second row, left panel:  $L_{e}-L_{\tau}$ with $\pi^{\pm}$ decay source. The second row, right panel: $L_{\mu}-L_{\tau}$ with $\pi^{\pm}$ decay source. The flavor ratio for $\mu^{\pm}$ damping is $\phi^{S}_{\alpha}=(0,1,0)$, and ratio for $\pi^{\pm}$ decay is $\phi^{S}_{\alpha}=(1/3,2/3,0)$. }
	\label{fig:3}
\end{figure}

\begin{table}
	\centering
	\caption{\label{Tab:2} The potential parameters in the models with $\mu^{\pm}$ damping source.}
	\begin{tabular}{c c c}
		\noalign{\smallskip}\hline
		\noalign{\smallskip}\hline
		
		~~Different models ~~~&~~~~~~~&~~~~[ Best fit value , Upper limit(95\% C.L.) ] [GeV] ~~\\
		\noalign{\smallskip}\hline
		~~\multirow{2}*{$L_{\alpha}-L_{\beta}$  symmetric model}
		~~~&~~~$V_{e\tau}$~~~~&~~~~$[~1.57*10^{-28},4.89*10^{-28}$~]~~\\
		~~~&~~~$V_{\mu\tau}$~~~~&~~~~$[~1.09*10^{-28},5.97*10^{-28}~]$~~\\
		\hline
		~~\multirow{2}*{$L_{\alpha}-L_{\beta}$ breaking model}
		~~~&~~~$V_{e\tau}$~~~~&~~~~$[~5.99*10^{-28},1.17*10^{-27}~]$~~\\	
		~~~&~~~$V_{\mu\tau}$~~~~&~~~~$[~2.13*10^{-28},6.01*10^{-28}~]$~~\\
		\hline
	\end{tabular}
\end{table}
\begin{table}
	\centering
	\caption{\label{Tab:3} The potential  parameters in the models with $\pi^{\pm}$ decay source.}
	\begin{tabular}{c c c}
		\noalign{\smallskip}\hline
		\noalign{\smallskip}\hline
		
		~~Different models ~~~&~~~~~~~&~~~~[ Best fit value , Upper limit(95\% C.L.) ] [GeV] ~~\\
		\noalign{\smallskip}\hline
		~~\multirow{2}*{$L_{\alpha}-L_{\beta}$ symmetric model}
		~~~&~~~$V_{e\tau}$~~~~&~~~~$[~8.27*10^{-27},1.86*10^{-26}~]$~~\\	
		~~~&~~~$V_{\mu\tau}$~~~~&~~~~$[~3.74*10^{-27},1.21*10^{-26}~]$~~\\
		\hline
		~~\multirow{2}*{$L_{\alpha}-L_{\beta}$ breaking model}
		~~~&~~~$V_{e\tau}$~~~~&~~~~$[~8.79*10^{-27},1.58*10^{-26}~]$~~\\
		~~~&~~~$V_{\mu\tau}$~~~~&~~~~$[~3.01*10^{-27},5.17*10^{-26}~]$~~\\
		\hline
	\end{tabular}
\end{table}

\section{Impacts of NSI on PeV neutrinos with and without $L_{\alpha}-L_{\beta}$ symmetry}
\subsection{Influences on the flavor ratio of high energy astrophysical neutrinos}
\begin{figure}\label{fig:4}
	\centering
	\includegraphics[width=0.45\textwidth]{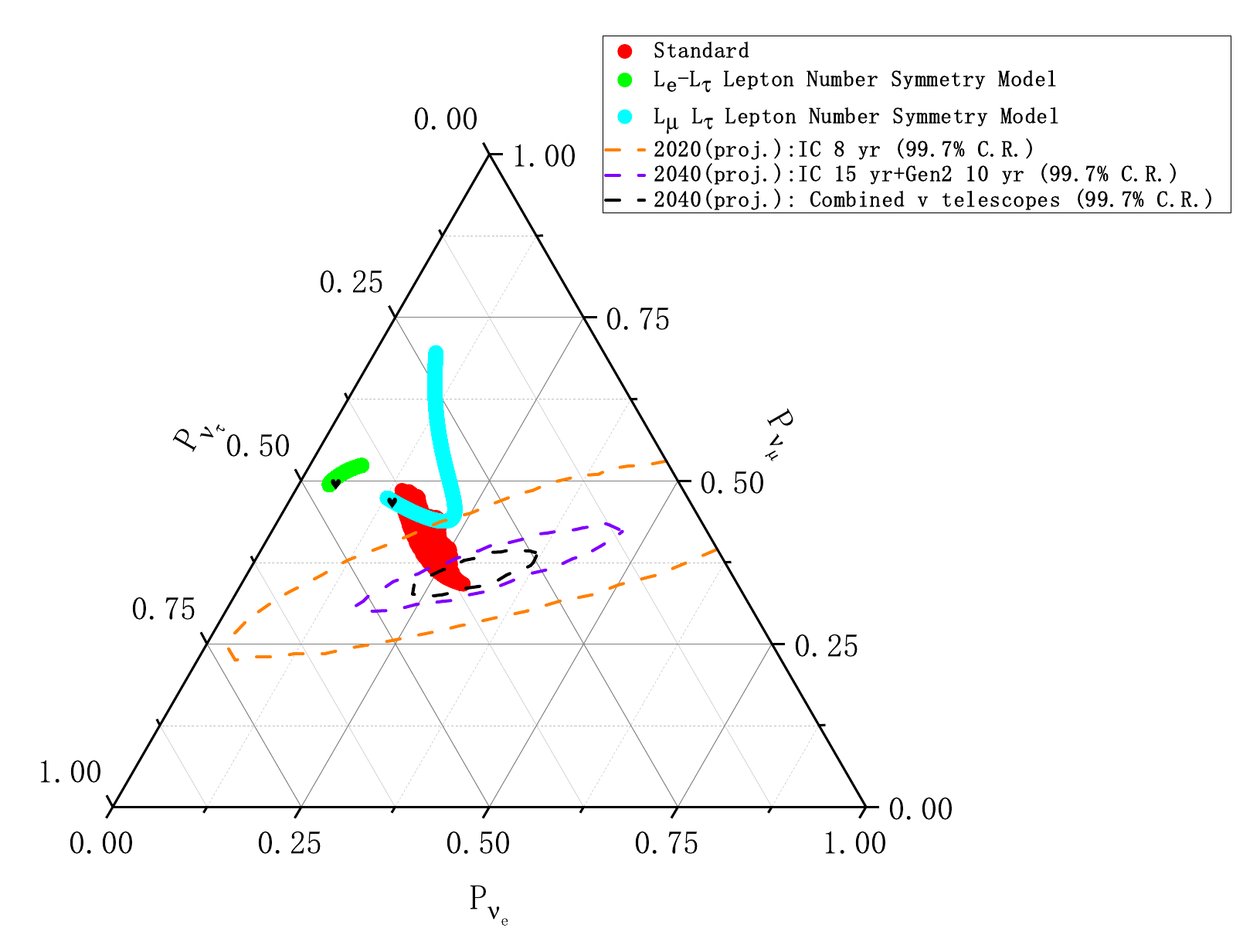}
	\includegraphics[width=0.45\textwidth]{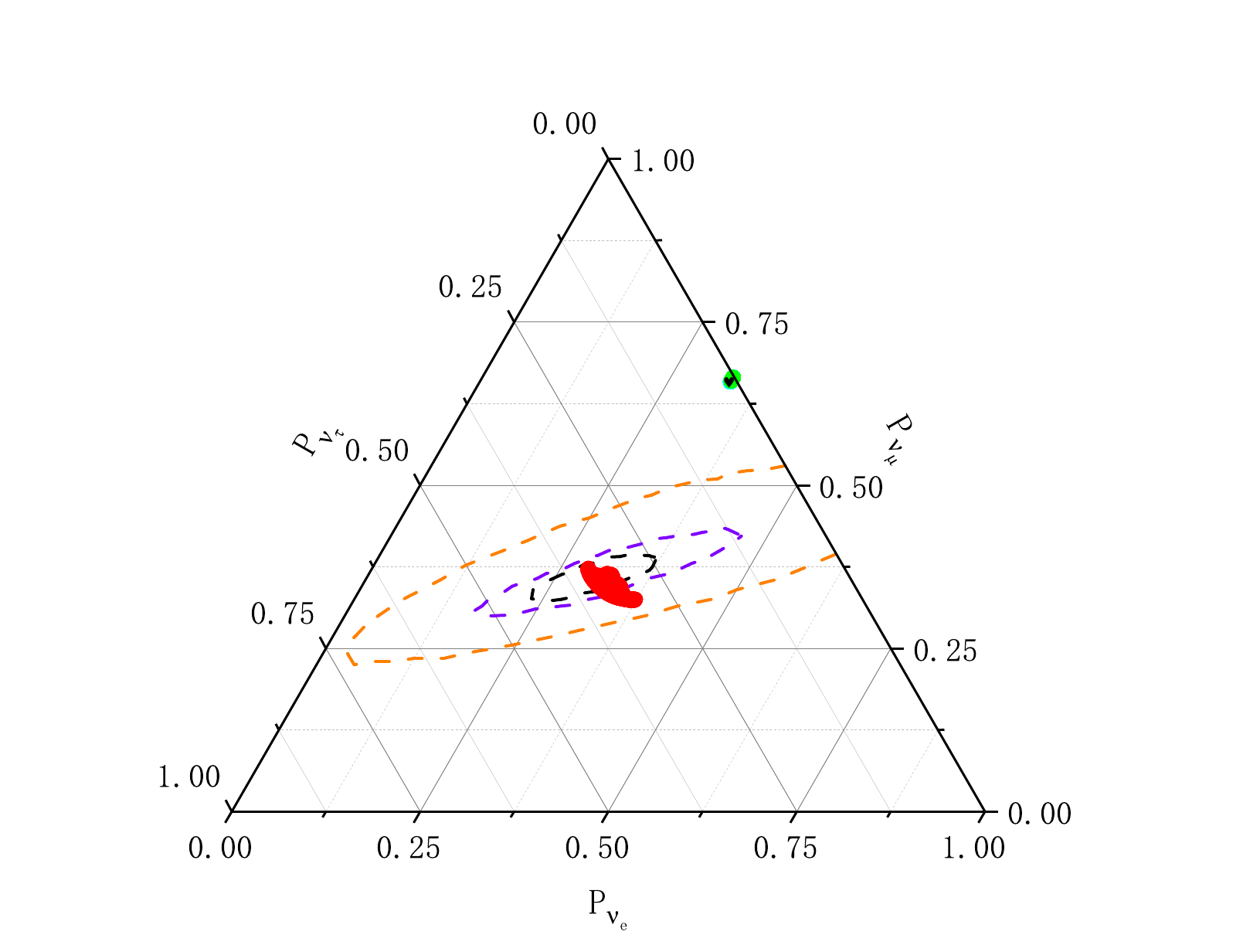}
	\caption{\label{fig:4} Ternary plot of the flavor ratio at Earth in the $L_{\alpha}-L_{\beta}$  symmetric models. The fixed neutrino energy is 1PeV, Neutrino oscillation parameters are taken in the $3\sigma$ allowed range of $NuFIT$ 5.0 for the global fit data(NO).  The left panel: $\mu^{\pm}$ damping source. Right panel: $\pi^{\pm}$ decay source. The orange dashed lines encompasses the 2020 $3\sigma$ C.R. with the $\pi$ decay source based on IceCube\cite{IceCube-Gen2:2020qha}.  The purple dashed line covers the 2040 $3\sigma$ C.R. based on IceCube and IceCube-Gen2\cite{IceCube-Gen2:2020qha}. The black dash lines  denotes the $3\sigma$ C.R. boundary based on the TeV-PeV neutrino telescopes available in 2040\cite{Song:2020nfh}. The black heart symbols represent the best-fit points.  }	
\end{figure}
\begin{figure}\label{fig:5}
	\centering
	\includegraphics[width=0.45\textwidth]{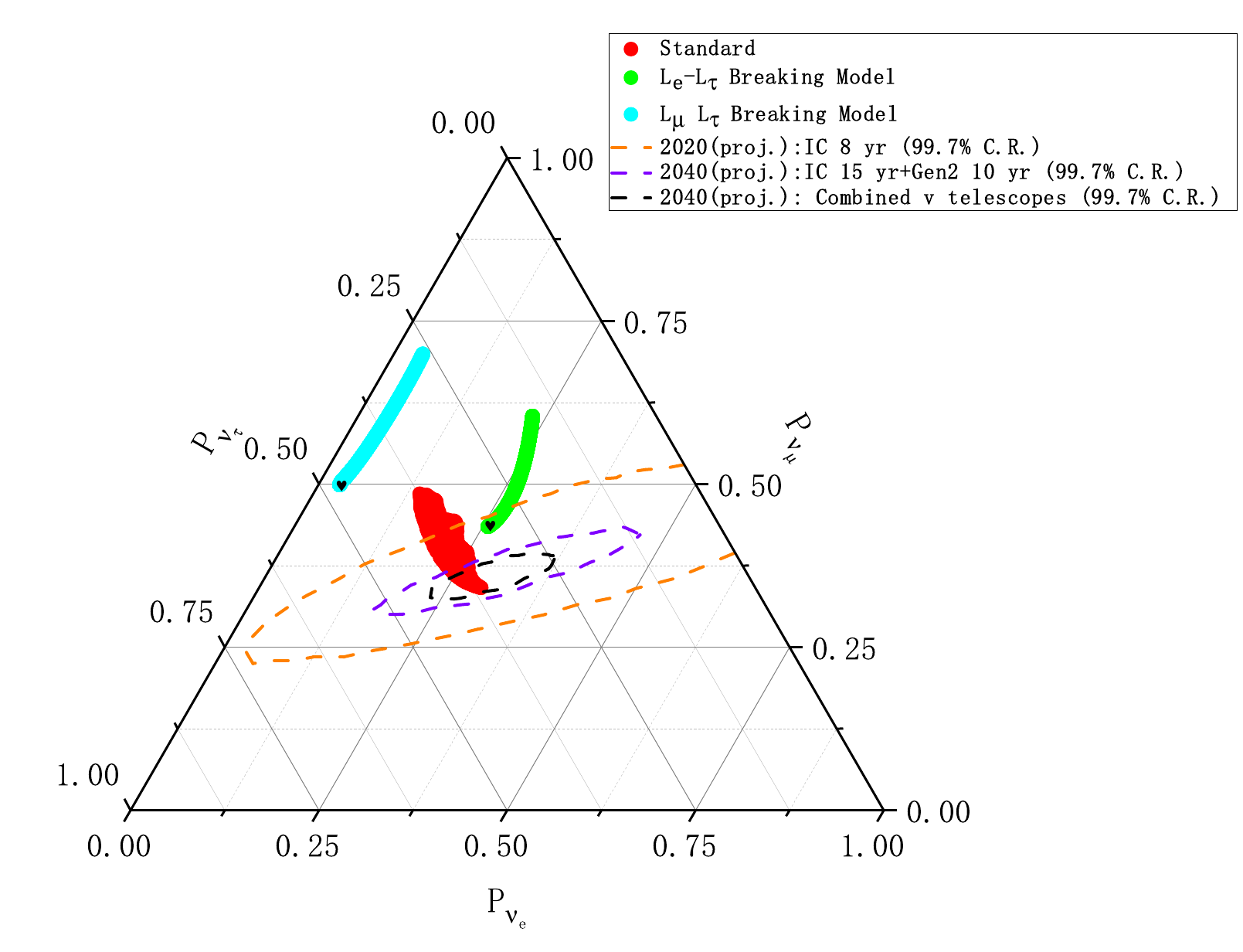}
	\includegraphics[width=0.45\textwidth]{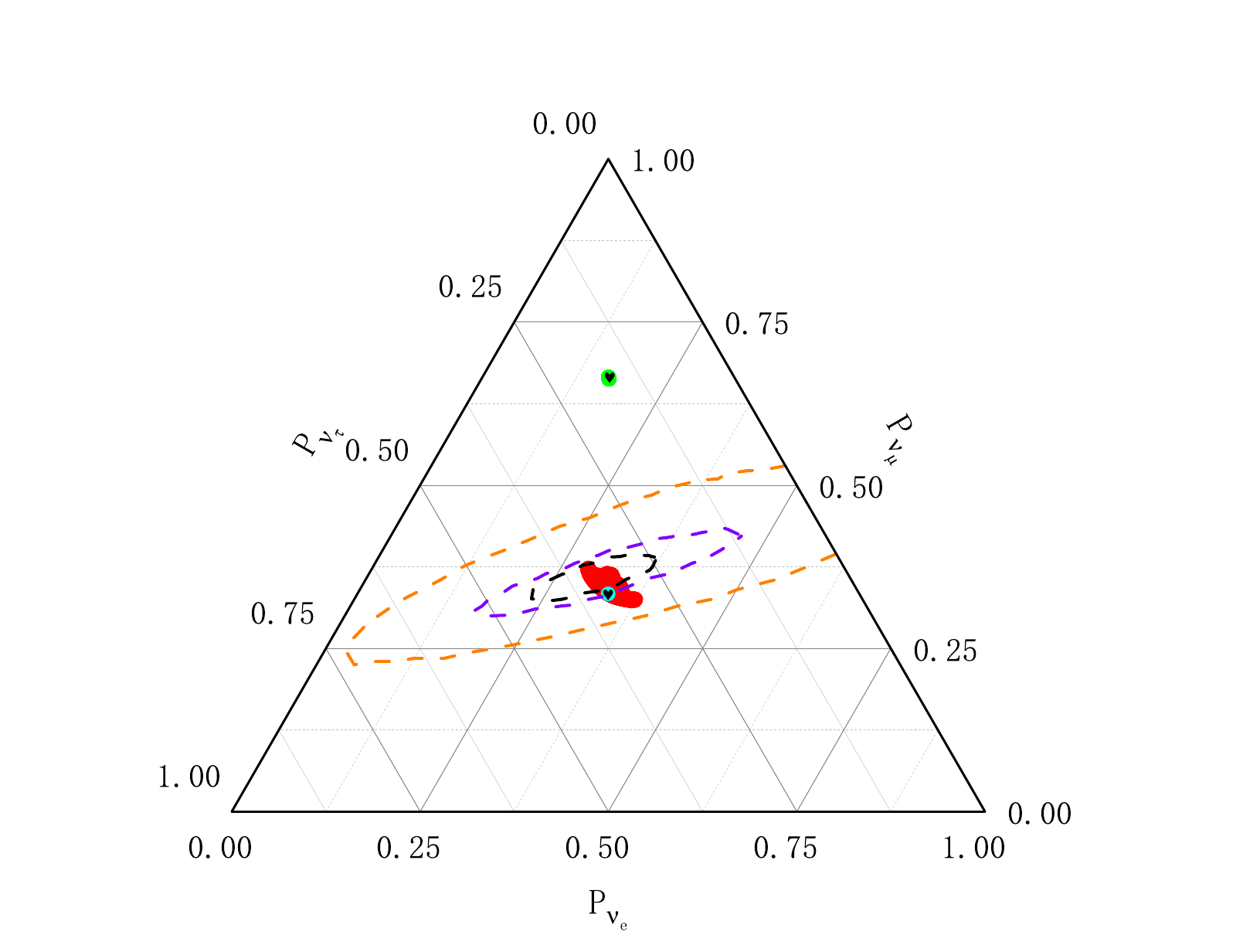}
	\caption{\label{fig:5} Ternary plot of the flavor ratio at Earth in the $L_{\alpha}-L_{\beta}$  breaking model. The left panel: $\mu^{\pm}$ damping source. Right panel: $\pi^{\pm}$ decay source.}
\end{figure}

In this section, the effects of new interactions on the detection of PeV neutrinos are examined. We first study the impacts of NSIs on the flavor composition of astrophysical neutrinos at Earth. As an illustrative example, the neutrino energy is taken 1 PeV, assuming NO leptonic mixing parameters. We consider two typical sources, namely $\mu^{\pm}$ damping source with $\phi^{S}_{\alpha}=(0,1,0)$, and $\pi^{\pm}$ decay source with $\phi^{S}_{\alpha}=(1/3,2/3,0)$. The potential parameter $V_{\alpha\beta}$ takes value in the ranges listed in Tab.\ref{Tab:2}-\ref{Tab:3}, namely from its best fit value to the 0.95 upper limit. Based on the given parameters, we show ternary plots of the flavor ratio at Earth (see Fig.\ref{fig:4}- \ref{fig:5}). For ease of comparison, the ternary plots also show the area from the standard flavor conversion scheme. The following observations can be obtained from these figures.

For $\mu^{\pm}$ damping sources, the predictions of both the  $L_{\alpha}-L_{\beta}$ symmetric and breaking models, deviate noticeably from the result of the standard model.
In contrast, for the models under the $\pi^{\pm}$ decay sources, the flavor ratio of PeV neutrinos is almost concentrated at special points. In particular, in the symmetric case (See right panel of Fig.\ref{fig:4})) the regions of $L_{e}-L_{\tau}$ and $L_{\mu}-L_{\tau}$ case overlap each other.

Let us give a simple comment on the difference. When the potential parameter $V_{\alpha\beta}$ reaches a threshold,
it would dominate the Hamilton and the flavor transition matrix could be fixed on  a decoupling  pattern ((see Eqs.\ref{eq:17}- \ref{eq:18}) in the following section).
Correspondingly, the flavor ratio at Earth could be given on a special point. Based on the ranges of parameter values in Tab.\ref{Tab:2}-\ref{Tab:3}, we can see that
the parameters with the $\pi^{\pm}$ decay source approximate the flavor decoupling thresholds both in the $L_{\alpha}-L_{\beta}$ symmetric and breaking models. Hence
the concentration of flavor ratio region appears.

\subsection{Influences on the energy spectrum of astrophysical neutrinos}
The NSI interactions bring a significant effect on the flavor conversion probability of high-energy astrophysical neutrinos, which can cause the variation of the flux of a special flavor at Earth. For the flux $\phi_{\overline{\nu}_{e}}$ produced from the $pp$ collision, we have
\begin{equation}\label{eq:16}
	\phi_{\overline{\nu}_{e}}(E)=\frac{1}{2}\phi_{\nu+\overline{\nu}}(E)\times(\sum_{\alpha=e,\mu,\tau}\overline{P}_{e\alpha}(E,V)~\phi^{S}_{\alpha}~).
\end{equation}
For $\phi_{\nu_{\tau}}$ and $\phi_{\overline{\nu}_{\tau}}$, the corresponding expression is obtained by the replacement $e\rightarrow\tau$ in $\overline{P}_{e\alpha}(E,V)$, with $\phi_{\overline{\nu}_{\tau}+\nu_{\tau}}=2\phi_{\nu_{\tau}}$.
Based on the  $\phi_{\nu+\overline{\nu}}(E)$ of single power law (eq.\ref{eq:12}) and the neutrino oscillation and potential parameters specified in Fig.\ref{fig:4}- \ref{fig:5}, we show the influences of NSI on the energy spectrum of $\overline{\nu}_{e}$ and $\nu_{\tau}$(see Fig.\ref{fig:6}-\ref{fig:9}).

\begin{figure}\label{fig:6}
	\centering
	\includegraphics[width=0.45\textwidth]{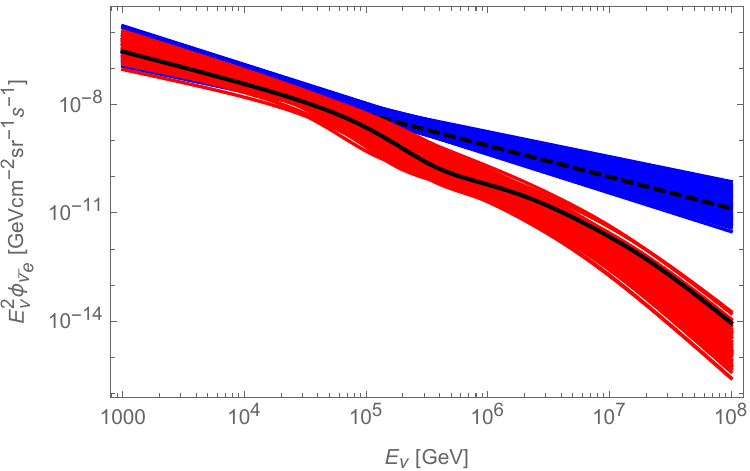}
	\hfill
	\includegraphics[width=0.45\textwidth]{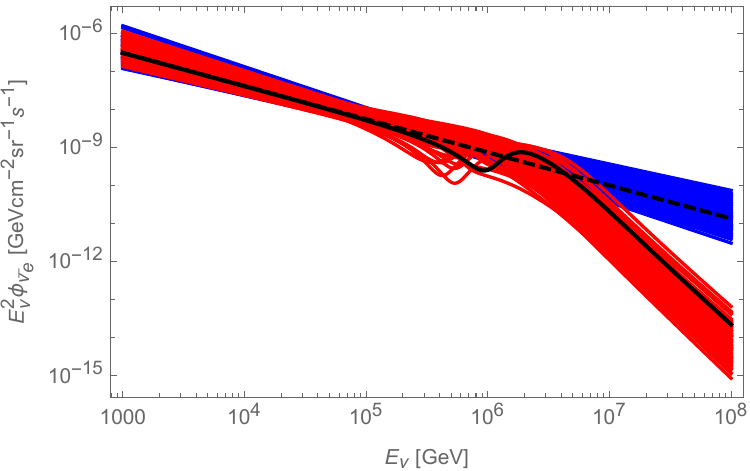}
	\hfill
	\includegraphics[width=0.45\textwidth]{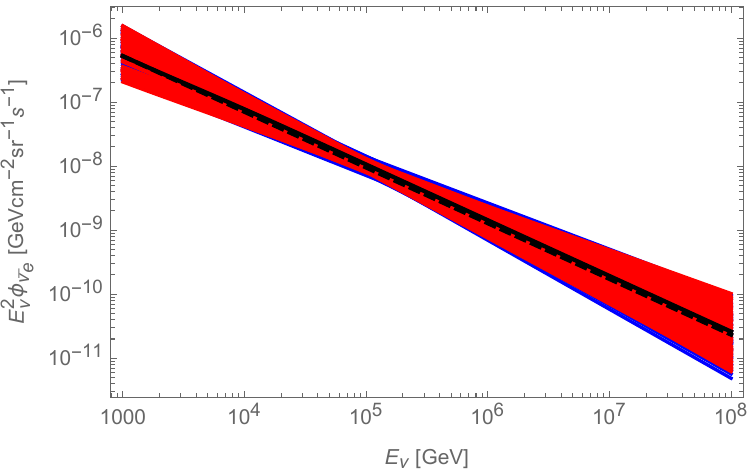}
	\hfill
	\includegraphics[width=0.45\textwidth]{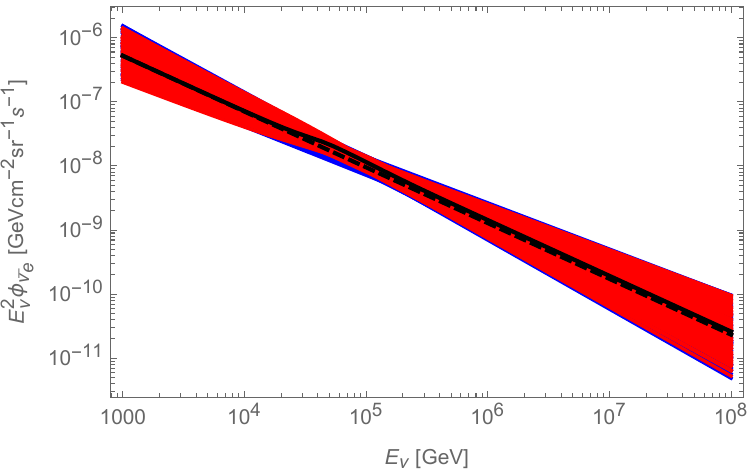}
	\hfill
	\caption{\label{fig:6} The energy spectrum of $\overline{\nu}_{e}$ in the $L_{\alpha}-L_{\beta}$ symmetric model. The first row, left panel: $L_{e}-L_{\tau}$ case with $\mu^{\pm}$ damping source, right panel: $L_{\mu}-L_{\tau}$ case with $\mu^{\pm}$ damping sources. The second row, left panel : $L_{e}-L_{\tau}$ case with $\pi^{\pm}$ decay source, right panel: $L_{\mu}-L_{\tau}$ case with $\pi^{\pm}$ decay sources. The ranges of the parameters of the energy spectrum $\phi_{\nu+\overline{\nu}}$ are taken  in the 68\%  C.L. in Tab.\ref{Tab:1}. The potential parameters $V_{\alpha\beta}$ are the same as the values given in Fig.\ref{fig:4}. The neutrinos oscillation parameters are taken in the $3\sigma$ allowed ranges of the global fit data of NO. Blue lines: arising from the standard transition matrix $\overline{P^{s}}$ without the long range potential effects. Red lines: from the matrix $\overline{P}$ including the long range potential effects. The black line: arising from the $\overline{P}$ with the best fit values of all the given parameters. The dashed black line: from $\overline{P^{s}}$ with the best fit values of the given parameters.}
\end{figure}
\begin{figure}
	\centering
	\includegraphics[width=0.45\textwidth]{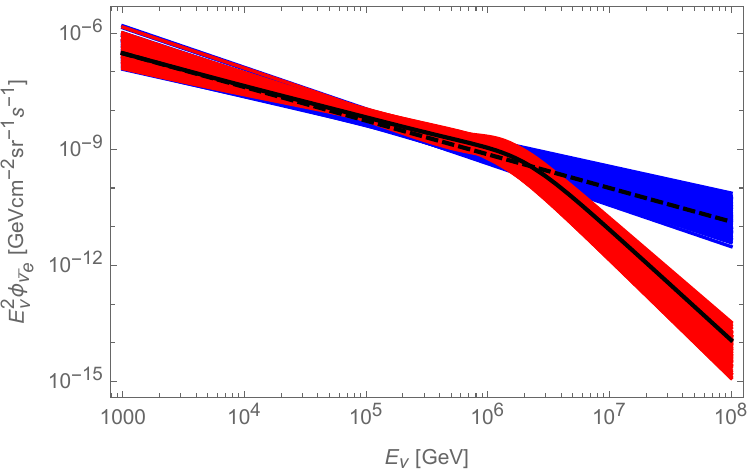}
	\hfill
	\includegraphics[width=0.45\textwidth]{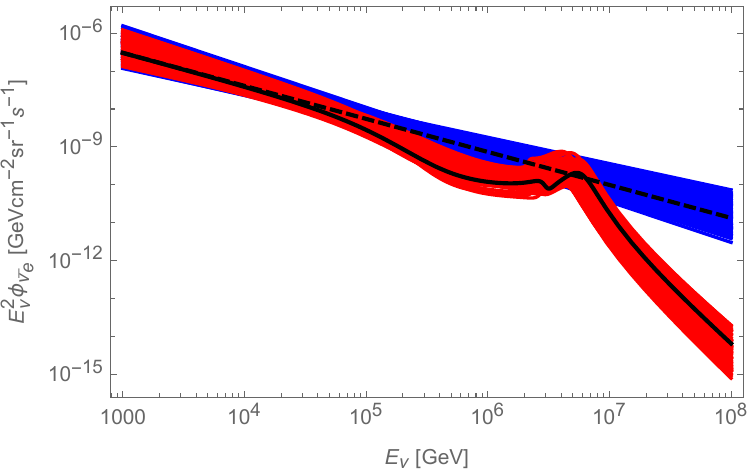}
	\hfill
	\includegraphics[width=0.45\textwidth]{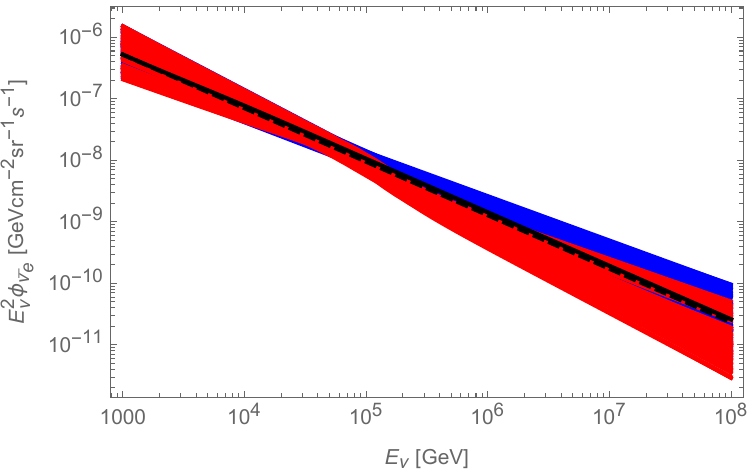}
	\hfill
	\includegraphics[width=0.45\textwidth]{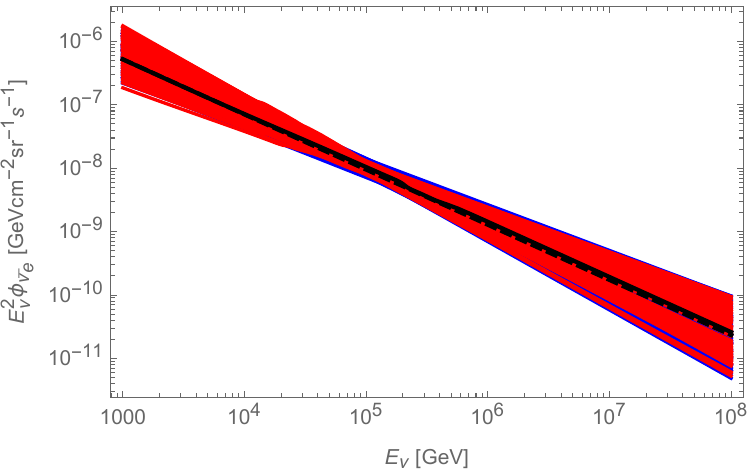}
	\hfill
	\caption{\label{fig:7} The energy spectrum of $\overline{\nu}_{e}$ in the symmetry breaking model. The first row, left panel: $L_{e}-L_{\tau}$ case with $\mu^{\pm}$ damping source, right panel: $L_{\mu}-L_{\tau}$ case with $\mu^{\pm}$ damping sources. The second row, left panel : $L_{e}-L_{\tau}$ case with $\pi^{\pm}$ decay source, right panel: $L_{\mu}-L_{\tau}$ case with $\pi^{\pm}$ decay sources. The potential parameters $V_{\alpha\beta}$ are the same as the values given in Fig.\ref{fig:5}. The conventions for other parameters, colors and lines are the same as those in Fig.\ref{fig:6}.}
\end{figure}
\begin{figure}
	\centering
	\includegraphics[width=0.45\textwidth]{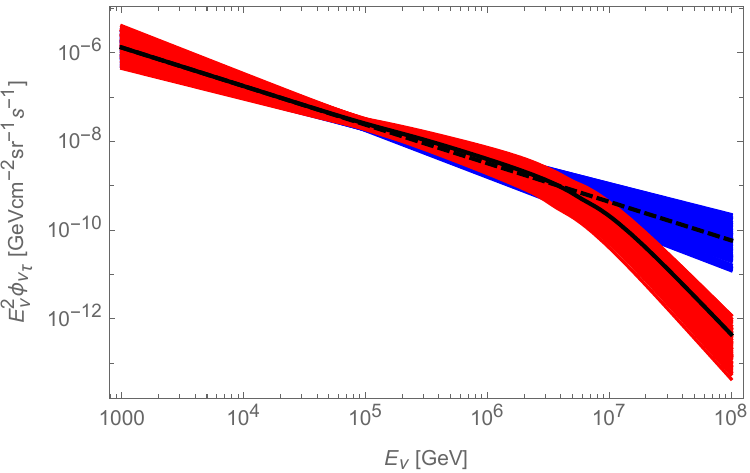}
	\hfill
	\includegraphics[width=0.45\textwidth]{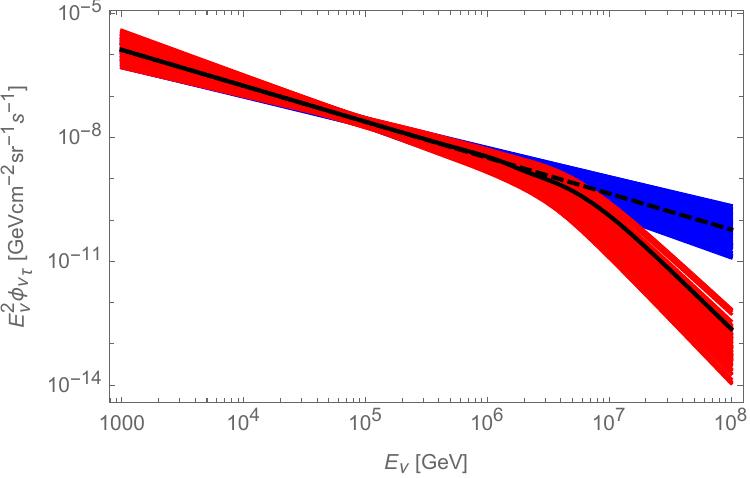}
	\hfill
	\includegraphics[width=0.45\textwidth]{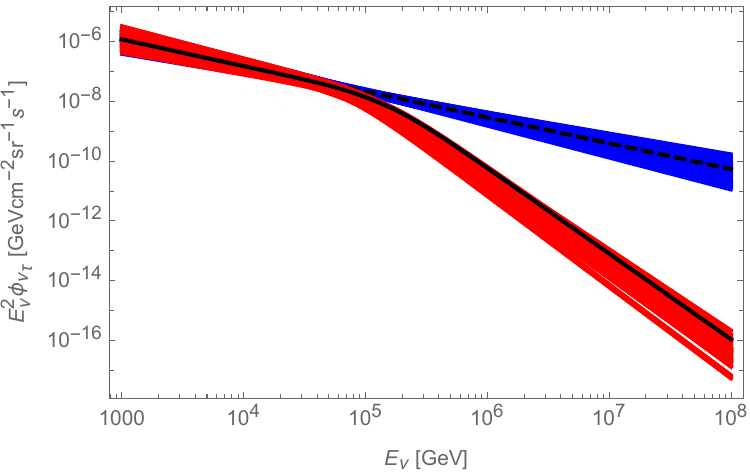}
	\hfill
	\includegraphics[width=0.45\textwidth]{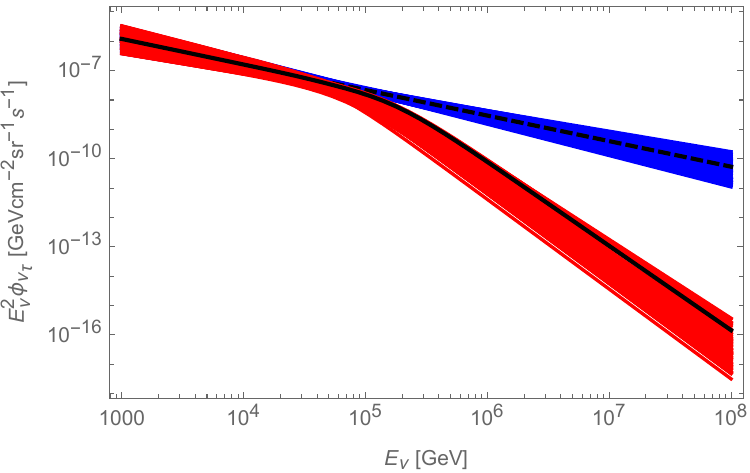}
	\hfill
	\caption{\label{fig:8} The energy spectrum of ${\nu}_{\tau}$ in the $L_{\alpha}-L_{\beta}$ symmetric model. The first row, left panel: $L_{e}-L_{\tau}$ case with $\mu^{\pm}$ damping source, right panel: $L_{\mu}-L_{\tau}$ case with $\mu^{\pm}$ damping sources. The second row, left panel : $L_{e}-L_{\tau}$ case with $\pi^{\pm}$ decay source, right panel: $L_{\mu}-L_{\tau}$ case with $\pi^{\pm}$ decay sources. The conventions for parameters, colors and lines are the same as those in Fig.\ref{fig:6}.}
\end{figure}
\begin{figure}
	\centering
	\includegraphics[width=0.45\textwidth]{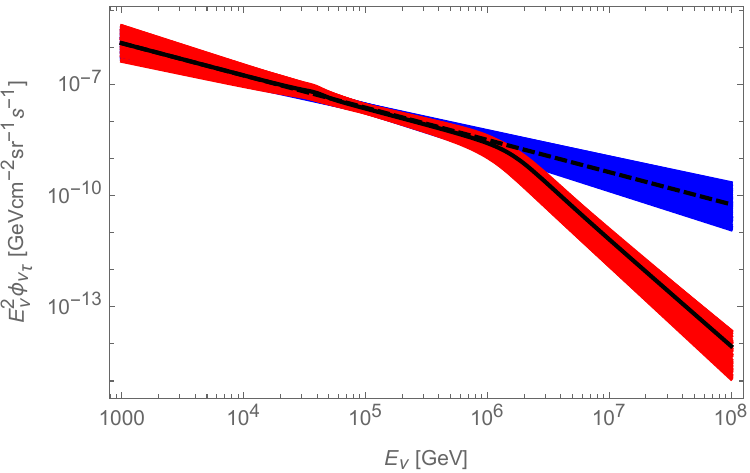}
	\hfill
	\includegraphics[width=0.45\textwidth]{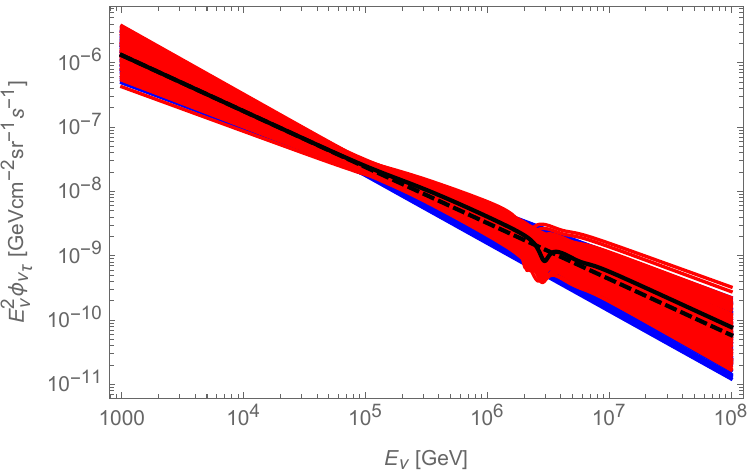}
	\hfill
	\includegraphics[width=0.45\textwidth]{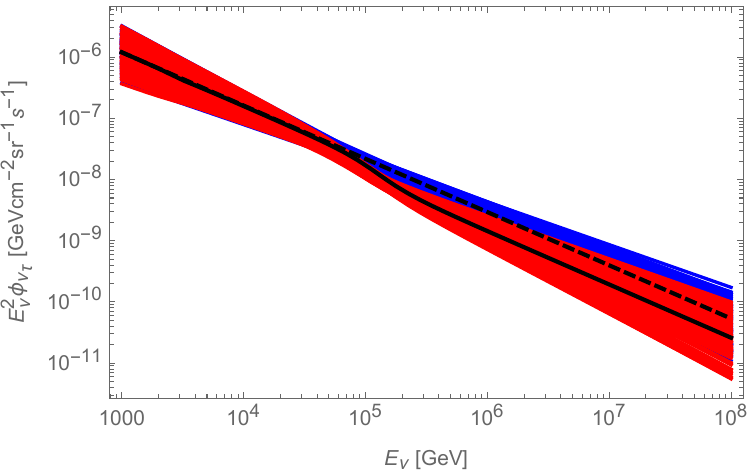}
	\hfill
	\includegraphics[width=0.45\textwidth]{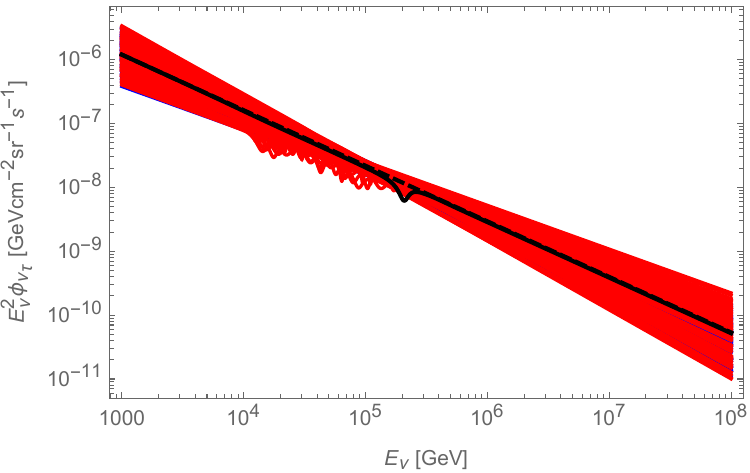}
	\hfill
	\caption{\label{fig:9} The energy spectrum of ${\nu}_{\tau}$ in the $L_{\alpha}-L_{\beta}$ breaking model. The first row, left panel: $L_{e}-L_{\tau}$ case with $\mu^{\pm}$ damping source, right panel: $L_{\mu}-L_{\tau}$ case with $\mu^{\pm}$ damping sources. The second row, left panel : $L_{e}-L_{\tau}$ case with $\pi^{\pm}$ decay source, right panel: $L_{\mu}-L_{\tau}$ case with $\pi^{\pm}$ decay sources. The conventions for parameters, colors and lines are the same as those in Fig.\ref{fig:7}.}
\end{figure}

As is seen in  the Fig.\ref{fig:6}-Fig.\ref{fig:7}, in the range $E_{\nu}>5$PeV the flux $\phi_{\overline{\nu}_{e}}(E)$ is apparently damped for neutrinos from the $\mu^{\pm}$ damping source both in the $L_{\alpha}-L_{\beta}$ symmetric and breaking models with the parameter $V_{\alpha\beta}$ around their best fit data.
 As a comparison, the damping of $\phi_{\overline{\nu}_{e}}(E)$ in the case of $\pi^{\pm}$ decay source is undermined by by the large uncertainty of the parameters of $\phi_{\nu+\overline{\nu}}$.

Let us give some comments on the suppressing effect on $\phi_{\overline{\nu}_{e}}$ from NSIs.  Since the factor $E^{-1}$ is in the Hamiltonian $H_{0}$, for PeV neutrinos in the $L_{\alpha}-L_{\beta}$ symmetric model, when the long range potential $V_{\alpha\beta}>6.29*10^{-27}$GeV, the flavor transition matrix approximates the unit matrix, i.e.,
\begin{equation}\label{eq:17}
		\overline{P}\simeq
			\left(
			\begin{array}{ccc}
				1&0&0\\
				0&1&0\\
				0&0&1
			\end{array}
			\right).
\end{equation}
Therefore, the flavor ratio of the neutrinos at Earth is nearly equal to the value at the source. In particular, for the $\mu^{\pm}$ damping source, $\phi_{\overline{\nu}_{e}}$
 becomes negligible at Earth.

In the breaking models, when the parameter $V_{\alpha\beta}>9*10^{-27}$GeV, the conversion matrix with $E_{\nu}>1$PeV also shows the characteristic of flavor decoupling, namely
\begin{equation}\label{eq:18}
	\overline{P}(V_{e\tau})\simeq
	\left(
	\begin{array}{ccc}
		\frac{1}{2}~&~0~&~\frac{1}{2}~\\
		0~&~1~&~0~\\
		\frac{1}{2}~&~0~&~\frac{1}{2}
	\end{array}
	\right);~~
	\overline{P}(V_{\mu\tau})\simeq
	\left(
	\begin{array}{ccc}
		1~&~0~&~0~\\
		0~&~\frac{1}{2}~&~ \frac{1}{2}~\\
		0~&~\frac{1}{2}~&~\frac{1}{2}
	\end{array}
	\right).
\end{equation}
Thus, the flux of $\nu_{e}$($\overline{\nu}_{e}$) at the PeV energy-scale  from the $\mu^{\pm}$ damping source is also noticeably reduced by the NSI.

 Because of the flavor decoupling from the $V_{\alpha\beta}$ dominated Hamiltonian, similar observations are  obtained for the flux $\phi_{\nu_{\tau}}$, see Fig.\ref{fig:8}-\ref{fig:9}. In this case, the suppressing of $\phi_{\nu_{\tau}}$ at PeV energies in the $L_{\alpha}-L_{\beta}$ symmetric model is remarkable for neutrinos from the $\pi^{\pm}$ decay source .

\subsection{Influences on the events of PeV neutrinos}
 Now we discuss the influence of NSI on the Glashow resonance and PeV $\nu_{\tau}(\overline{\nu}_{\tau})$ events with and without a $L_{\alpha}-L_{\beta}$ symmetry. As is known, Glashow resonance events provide a window to explore new physics at the PeV energy-scale. For the description of the intensity of the resonance events, V. Barger and et al. proposed a measure  in terms of the ratio of resonant events to non-resonant continuous events, namely\cite{Barger:2014iua}
\begin{equation}\label{eq:19}
	\begin{aligned}
		\frac{N_{Res}}{N_{non-Res}(E>E^{min})}&=\frac{10\pi}{18}(\frac{\Gamma_{W}}{M_{W}})(\frac{\sigma^{peak}_{Res}}{\sigma^{CC}_{\nu N}(E=6.3\rm PeV)})\frac{(\alpha-1.4)(\frac{E^{min}}{6.3\rm PeV})^{\alpha-1.4}}{[1-(\frac{E^{min}}{E^{max}})^{(\alpha-1.4)}]}\times [r_{e}]_{E=6.3\rm PeV} \\
		&=11\times\frac{(\alpha-1.4)(\frac{E^{min}}{6.3\rm PeV})^{\alpha-1.4}}{[1-(\frac{E^{min}}{E^{max}})^{(\alpha-1.4)}]}\times [r_{e}]_{E=6.3\rm PeV}~.
	\end{aligned}
\end{equation}
Here $r_{e}$ is fraction of $\overline{\nu}_{e}$ at Earth, the spectral index $\alpha$ is related to the neutrino acceleration mechanism. We let $\alpha$ be equal to 2 \cite{Anchordoqui:2013qsi,Anchordoqui:2013dnh}. Using the best fit data of the netrino oscillation parameters (NO), the potential parameter $V_{\alpha\beta}$ and the diffuse flux $\phi_{\nu+\overline{\nu}}$, the ratio of the Glashow resonance event is calculated near the resonance energy (see Tab.\ref{Tab:4}-\ref{Tab:5}). For comparison purposes, the results obtained from the standard flavor conversion matrix are also listed in the tables.
\begin{table}\caption{\label{Tab:4} The ratio of resonance event to nonresonant event with and without the NSI effect in the $L_{\alpha}-L_{\beta}$ symmetric model for the $\mu^{\pm}$ damped source and $\pi^{\pm}$ decay source. Here $\alpha~=~2$, $E^{min}$~=~1, 2, 3, 4, 5 PeV, $E^{max}~=~\infty$, $r_{e}$ takes the value $ [r_{e}]_{E~=~6.3\rm PeV}$}
	\centering
	\begin{tabular}{c c c c c c c c}
		\hline
		$E^{min}(\rm PeV)$&~~~&1&2&3&4&5&~~~$r_{e}$\\
		\hline
		\multirow{3}*{$(0~~1~~0)^{T}$}
		&$V_{e\tau}$&$6.3*10^{-3}$&$9.6*10^{-3}$&$1.2*10^{-2}$&$1.4*10^{-2}$&$1.6*10^{-2}$&$2.9*10^{-3}$\\
		&$V_{\mu\tau}$&$9.7*10^{-2}$&0.14&0.18&0.22&0.25&$4.4*10^{-2}$\\
		&SM&0.18 &0.28 &0.36 &0.43 & 0.49 &$8.6*10^{-2}$  \\
		\hline
		\multirow{3}*{$(1/3~~2/3~~0)^{T}$}
		&$V_{e\tau}$&0.36&0.55&0.70&0.83&0.95&0.16\\	
		&$V_{\mu\tau}$&0.36&0.55&0.70&0.83&0.95&0.16\\
		&SM& 0.32 &0.49&0.63&0.75&0.85&0.14\\
		\hline
	\end{tabular}
\end{table}
\begin{table}\caption{\label{Tab:5}	The ratio of resonance event  to nonresonant event  with and without the NSI effect in the $L_{\alpha}-L_{\beta}$ breaking model.}
	\centering
	\begin{tabular}{c c c c c c c c}
		\hline
		$E^{min}(\rm PeV)$&~~~ &1&2&3&4&5&~~~$r_{e}$\\
		\hline
		\multirow{3}*{$(0~~1~~0)^{T}$}
		&$V_{e\tau}$&$3.8*10^{-2}$&$5.7*10^{-2}$&$7.3*10^{-2}$&$8.7*10^{-2}$&0.10&$1.7*10^{-2}$\\
		&$V_{\mu\tau}$&0.20&0.30&0.38&0.46&0.52&$9.1*10^{-2}$\\
	    &SM&0.18&0.28&0.36&0.43&0.49&$8.6*10^{-2}$\\
		\hline
		\multirow{3}*{$(1/3~~2/3~~0)^{T}$}
		&$V_{e\tau}$&0.18&0.27&0.35&0.41&0.47&$8.3*10^{-2}$\\		
		&$V_{\mu\tau}$&0.36&0.55&0.70&0.83&0.95&0.16\\		
		&SM&0.32&0.49&0.63&0.75&0.85&0.14\\
		\hline
	\end{tabular}
\end{table}

We can see that the Glashow resonant event ratio is obviously decreased by the NSI for neutrinos from  $\mu^{\pm}$ damping source. At the best fit value of $V_{e\tau}$,
the resonant event ratio in the symmetric model is around 30 times smaller than the value in SM scheme, while in the symmetry breaking model the ratio is near 5 times smaller than the standard result. Similar to the discussions on the flavor ratio in the previous section, the difference results from the Hamiltonian under the two models, which affects the flavor transition matrix. Using the best fit value of the $V_{e\tau}$ parameter, the flavor conversion matrices of the two models are obtained as follows:
\begin{equation}\label{eq:20}
	\overline{P}^{sm}(V_{e\tau})\simeq
	\left(
	\begin{array}{ccc}
		0.98~&~0.0058~&~0.0048~\\
		0.0058~&~0.65~&~0.34~\\
		0.0048~&~0.34~&~0.65
	\end{array}
	\right),~~
	\overline{P}^{bm}(V_{e\tau})\simeq
	\left(
	\begin{array}{ccc}
		0.48~&~0.034~&~0.48~\\
		0.034~&~0.94~&~ 0.016~\\
		0.48~&~0.016~&~0.49
	\end{array}
	\right),
\end{equation}
where $\overline{P}^{sm}(V_{e\tau})$ and $\overline{P}^{bm}(V_{e\tau})$ represent the matrix  under the symmetric and symmetry breaking model respectively.
We can see that the former matrix with the $\mu^{\pm}$ damping source can reduce the fraction $r_{e}$ to the order $10^{-3}$, while the latter gives $r_{e}$ of order $10^{-2}$. The damping of the Glashow resonance event ratio is  proportional to the damping of  $r_{e}$. Thus, the difference takes place.
Furthermore, we note that the ratio is sensitive to the parameter $V_{e\tau}$. Under the 95\% upper limit, the ratio can be damped to $10^{-3}$ times the magnitude of the SM result. In contrast, the ratio is moderately reduced by the NSI for neutrinos from  $\pi^{\pm}$ decay source. These observations are consistent with those read from Fig.\ref{fig:6}-\ref{fig:7}.

In a similar way, we examine suppressing effects of NSIs on $\nu_{\tau}$ events. The number of $\nu_{\tau}$ events is calculated as follow
\begin{equation}\label{eq:20}
	N_{k}(V_{\alpha\beta})=4\pi T\int_{k}\Phi_{\nu_{\tau}+\overline{\nu}_{\tau}}(E,V_{\alpha\beta})A_{\tau,k}(E)dE.
\end{equation}
where T=4318 days is the data collection time IceCube Observatory. The parameter $V_{\alpha\beta}$ is at the best fit value. We compare the results in the NSI models with  those in SM method(see Tab.\ref{Tab:6}-\ref{Tab:7}). In this case, the damping of the events is notable for neutrinos from $\pi^{\pm}$ decay sources in the $L_{\alpha}-L_{\beta}$ symmetric model. Thus, the symmetric NSI, making the detection of $\nu_{\tau}$ events at PeV  from the source impossible, would be stringently constrained by observations in the near future.

\begin{table}\caption{\label{Tab:6} The influence of NSI on the number of $\nu_{\tau}$ events in the $L_{\alpha}-L_{\beta}$ symmetric model.}
	\begin{tabular}{c c c c c c c }
		\hline
		$E(\rm PeV)$&~~~~&~~~~2.3~-~3.3~~~~&~~~~2.3~-~4~~~~&~~~~1.9~-~10~~~~\\
		\hline
		\multirow{3}*{$(0~~1~~0)^{T}$}
		&$V_{e\tau}$&0.27&0.35&0.72\\
		&$V_{\mu\tau}$&0.19&0.25&0.35\\
		&SM&0.23&0.31&0.68\\
		\hline
		\multirow{3}*{$(1/3~~2/3~~0)^{T}$}
		&$V_{e\tau}$&$7.2\times10^{-4}$&$8.6\times10^{-4}$&$1.8\times10^{-3}$\\
		&$V_{\mu\tau}$&$9.6\times10^{-4}$&$1.1\times10^{-3}$&$2.5\times10^{-3}$\\
		&SM&0.21&0.28&0.62\\
		\hline
	\end{tabular}
\end{table}

\begin{table}\caption{\label{Tab:7} The influence of NSI on the number of $\nu_{\tau}$ events in the $L_{\alpha}-L_{\beta}$ breaking model.}
	\begin{tabular}{c c c c c c c }
		\hline
		$E(\rm PeV)$&~~~~&~~~~2.3~-~3.3~~~~&~~~~2.3~-~4~~~~&~~~~1.9~-~10~~~~\\
		\hline
		\multirow{3}*{$(0~~1~~0)^{T}$}
		&$V_{e\tau}$&0.067&0.079&0.18\\
		&$V_{\mu\tau}$&0.22&0.30&0.72\\
		&SM&0.23&0.31&0.68\\
		\hline
		\multirow{3}*{$(1/3~~2/3~~0)^{T}$}
		&$V_{e\tau}$&0.10&0.14&0.30\\
		&$V_{\mu\tau}$&0.21&0.28&0.59\\
		&SM&0.21&0.28&0.62\\
		\hline
	\end{tabular}
\end{table}

\section{conclusion}
We examined the effects of NSI with and without a $L_{\alpha}-L_{\beta}$ symmetry on the HAN events at PeV energies. The constraints on potential parameters of NSI are
obtained with the updated IceCube shower data. On the bases the given parameters, the impacts of NSI on the flavor ratio of HANs at Earth, the flux $\phi_{\nu}$, and expected events number at IceCube Observatory are shown. In general, the considered NSI with a $L_{\alpha}-L_{\beta}$ symmetry can notably reduce the Glashow resonance events for neutrinos from  $\mu^{\pm}$ damping source and the $\nu_{\tau}$ events for neutrinos from  $\pi^{\pm}$ decay source. Considering the upcoming neutrino observatories, the observations of PeV
events may give more stringent constraints on the NSIs with a $L_{\alpha}-L_{\beta}$ symmetry.

\vspace{0.08cm}

\acknowledgments
This work is supported by the National Natural Science Foundation of China under grant No. 12065007, the Research Foundation of Guilin University of Technology under grant No. GUTQDJJ2018103.

\bibliography{refs3}

\begin{thebibliography}{62}%
\makeatletter
\providecommand \@ifxundefined [1]{%
 \@ifx{#1\undefined}
}%
\providecommand \@ifnum [1]{%
 \ifnum #1\expandafter \@firstoftwo
 \else \expandafter \@secondoftwo
 \fi
}%
\providecommand \@ifx [1]{%
 \ifx #1\expandafter \@firstoftwo
 \else \expandafter \@secondoftwo
 \fi
}%
\providecommand \natexlab [1]{#1}%
\providecommand \enquote  [1]{``#1''}%
\providecommand \bibnamefont  [1]{#1}%
\providecommand \bibfnamefont [1]{#1}%
\providecommand \citenamefont [1]{#1}%
\providecommand \href@noop [0]{\@secondoftwo}%
\providecommand \href [0]{\begingroup \@sanitize@url \@href}%
\providecommand \@href[1]{\@@startlink{#1}\@@href}%
\providecommand \@@href[1]{\endgroup#1\@@endlink}%
\providecommand \@sanitize@url [0]{\catcode `\\12\catcode `\$12\catcode
  `\&12\catcode `\#12\catcode `\^12\catcode `\_12\catcode `\%12\relax}%
\providecommand \@@startlink[1]{}%
\providecommand \@@endlink[0]{}%
\providecommand \url  [0]{\begingroup\@sanitize@url \@url }%
\providecommand \@url [1]{\endgroup\@href {#1}{\urlprefix }}%
\providecommand \urlprefix  [0]{URL }%
\providecommand \Eprint [0]{\href }%
\providecommand \doibase [0]{http://dx.doi.org/}%
\providecommand \selectlanguage [0]{\@gobble}%
\providecommand \bibinfo  [0]{\@secondoftwo}%
\providecommand \bibfield  [0]{\@secondoftwo}%
\providecommand \translation [1]{[#1]}%
\providecommand \BibitemOpen [0]{}%
\providecommand \bibitemStop [0]{}%
\providecommand \bibitemNoStop [0]{.\EOS\space}%
\providecommand \EOS [0]{\spacefactor3000\relax}%
\providecommand \BibitemShut  [1]{\csname bibitem#1\endcsname}%
\let\auto@bib@innerbib\@empty
%</preamble>
\bibitem [{\citenamefont {Aartsen}\ \emph
  {et~al.}(2013{\natexlab{a}})\citenamefont {Aartsen} \emph
  {et~al.}}]{IceCube:2013low}%
  \BibitemOpen
  \bibfield  {author} {\bibinfo {author} {\bibfnamefont {M.~G.}\ \bibnamefont
  {Aartsen}} \emph {et~al.} (\bibinfo {collaboration} {IceCube}),\ }\href
  {\doibase 10.1126/science.1242856} {\bibfield  {journal} {\bibinfo  {journal}
  {Science}\ }\textbf {\bibinfo {volume} {342}},\ \bibinfo {pages} {1242856}
  (\bibinfo {year} {2013}{\natexlab{a}})},\ \Eprint
  {http://arxiv.org/abs/1311.5238} {arXiv:1311.5238 [astro-ph.HE]} \BibitemShut
  {NoStop}%
\bibitem [{\citenamefont {Aartsen}\ \emph
  {et~al.}(2013{\natexlab{b}})\citenamefont {Aartsen} \emph
  {et~al.}}]{IceCube:2013cdw}%
  \BibitemOpen
  \bibfield  {author} {\bibinfo {author} {\bibfnamefont {M.~G.}\ \bibnamefont
  {Aartsen}} \emph {et~al.} (\bibinfo {collaboration} {IceCube}),\ }\href
  {\doibase 10.1103/PhysRevLett.111.021103} {\bibfield  {journal} {\bibinfo
  {journal} {Phys. Rev. Lett.}\ }\textbf {\bibinfo {volume} {111}},\ \bibinfo
  {pages} {021103} (\bibinfo {year} {2013}{\natexlab{b}})},\ \Eprint
  {http://arxiv.org/abs/1304.5356} {arXiv:1304.5356 [astro-ph.HE]} \BibitemShut
  {NoStop}%
\bibitem [{\citenamefont {Aartsen}\ \emph {et~al.}(2014)\citenamefont {Aartsen}
  \emph {et~al.}}]{IceCube:2014stg}%
  \BibitemOpen
  \bibfield  {author} {\bibinfo {author} {\bibfnamefont {M.~G.}\ \bibnamefont
  {Aartsen}} \emph {et~al.} (\bibinfo {collaboration} {IceCube}),\ }\href
  {\doibase 10.1103/PhysRevLett.113.101101} {\bibfield  {journal} {\bibinfo
  {journal} {Phys. Rev. Lett.}\ }\textbf {\bibinfo {volume} {113}},\ \bibinfo
  {pages} {101101} (\bibinfo {year} {2014})},\ \Eprint
  {http://arxiv.org/abs/1405.5303} {arXiv:1405.5303 [astro-ph.HE]} \BibitemShut
  {NoStop}%
\bibitem [{\citenamefont {Aartsen}\ \emph {et~al.}(2016)\citenamefont {Aartsen}
  \emph {et~al.}}]{IceCube:2016umi}%
  \BibitemOpen
  \bibfield  {author} {\bibinfo {author} {\bibfnamefont {M.~G.}\ \bibnamefont
  {Aartsen}} \emph {et~al.} (\bibinfo {collaboration} {IceCube}),\ }\href
  {\doibase 10.3847/0004-637X/833/1/3} {\bibfield  {journal} {\bibinfo
  {journal} {Astrophys. J.}\ }\textbf {\bibinfo {volume} {833}},\ \bibinfo
  {pages} {3} (\bibinfo {year} {2016})},\ \Eprint
  {http://arxiv.org/abs/1607.08006} {arXiv:1607.08006 [astro-ph.HE]}
  \BibitemShut {NoStop}%
\bibitem [{\citenamefont {Abbasi}\ \emph {et~al.}(2022)\citenamefont {Abbasi}
  \emph {et~al.}}]{IceCube:2021uhz}%
  \BibitemOpen
  \bibfield  {author} {\bibinfo {author} {\bibfnamefont {R.}~\bibnamefont
  {Abbasi}} \emph {et~al.} (\bibinfo {collaboration} {IceCube}),\ }\href
  {\doibase 10.3847/1538-4357/ac4d29} {\bibfield  {journal} {\bibinfo
  {journal} {Astrophys. J.}\ }\textbf {\bibinfo {volume} {928}},\ \bibinfo
  {pages} {50} (\bibinfo {year} {2022})},\ \Eprint
  {http://arxiv.org/abs/2111.10299} {arXiv:2111.10299 [astro-ph.HE]}
  \BibitemShut {NoStop}%
\bibitem [{\citenamefont {Colladay}\ and\ \citenamefont
  {Kostelecky}(1998)}]{Colladay:1998fq}%
  \BibitemOpen
  \bibfield  {author} {\bibinfo {author} {\bibfnamefont {D.}~\bibnamefont
  {Colladay}}\ and\ \bibinfo {author} {\bibfnamefont {V.~A.}\ \bibnamefont
  {Kostelecky}},\ }\href@noop {} {\bibfield  {journal} {\bibinfo  {journal}
  {Phys. Rev. D}\ }\textbf {\bibinfo {volume} {58}},\ \bibinfo {pages} {116002}
  (\bibinfo {year} {1998})},\ \Eprint {http://arxiv.org/abs/hep-ph/9809521}
  {arXiv:hep-ph/9809521} \BibitemShut {NoStop}%
\bibitem [{\citenamefont {Coleman}\ and\ \citenamefont
  {Glashow}(1999)}]{Coleman:1998ti}%
  \BibitemOpen
  \bibfield  {author} {\bibinfo {author} {\bibfnamefont {S.~R.}\ \bibnamefont
  {Coleman}}\ and\ \bibinfo {author} {\bibfnamefont {S.~L.}\ \bibnamefont
  {Glashow}},\ }\href@noop {} {\bibfield  {journal} {\bibinfo  {journal} {Phys.
  Rev. D}\ }\textbf {\bibinfo {volume} {59}},\ \bibinfo {pages} {116008}
  (\bibinfo {year} {1999})},\ \Eprint {http://arxiv.org/abs/hep-ph/9812418}
  {arXiv:hep-ph/9812418} \BibitemShut {NoStop}%
\bibitem [{\citenamefont {Kostelecky}\ and\ \citenamefont
  {Mewes}(2004)}]{Kostelecky:2003cr}%
  \BibitemOpen
  \bibfield  {author} {\bibinfo {author} {\bibfnamefont {V.~A.}\ \bibnamefont
  {Kostelecky}}\ and\ \bibinfo {author} {\bibfnamefont {M.}~\bibnamefont
  {Mewes}},\ }\href@noop {} {\bibfield  {journal} {\bibinfo  {journal} {Phys.
  Rev. D}\ }\textbf {\bibinfo {volume} {69}},\ \bibinfo {pages} {016005}
  (\bibinfo {year} {2004})},\ \Eprint {http://arxiv.org/abs/hep-ph/0309025}
  {arXiv:hep-ph/0309025} \BibitemShut {NoStop}%
\bibitem [{\citenamefont {Roberts}(2021)}]{Roberts:2021vsi}%
  \BibitemOpen
  \bibfield  {author} {\bibinfo {author} {\bibfnamefont {A.}~\bibnamefont
  {Roberts}},\ }\href@noop {} {\bibfield  {journal} {\bibinfo  {journal}
  {Galaxies}\ }\textbf {\bibinfo {volume} {9}},\ \bibinfo {pages} {47}
  (\bibinfo {year} {2021})}\BibitemShut {NoStop}%
\bibitem [{\citenamefont {Zhang}\ and\ \citenamefont
  {Yang}(2022)}]{Zhang:2022svg}%
  \BibitemOpen
  \bibfield  {author} {\bibinfo {author} {\bibfnamefont {H.}~\bibnamefont
  {Zhang}}\ and\ \bibinfo {author} {\bibfnamefont {L.}~\bibnamefont {Yang}},\
  }\href@noop {} {\bibfield  {journal} {\bibinfo  {journal} {Universe}\
  }\textbf {\bibinfo {volume} {8}},\ \bibinfo {pages} {260} (\bibinfo {year}
  {2022})}\BibitemShut {NoStop}%
\bibitem [{\citenamefont {Beacom}\ \emph {et~al.}(2003)\citenamefont {Beacom},
  \citenamefont {Bell}, \citenamefont {Hooper}, \citenamefont {Pakvasa},\ and\
  \citenamefont {Weiler}}]{Beacom:2002vi}%
  \BibitemOpen
  \bibfield  {author} {\bibinfo {author} {\bibfnamefont {J.~F.}\ \bibnamefont
  {Beacom}}, \bibinfo {author} {\bibfnamefont {N.~F.}\ \bibnamefont {Bell}},
  \bibinfo {author} {\bibfnamefont {D.}~\bibnamefont {Hooper}}, \bibinfo
  {author} {\bibfnamefont {S.}~\bibnamefont {Pakvasa}}, \ and\ \bibinfo
  {author} {\bibfnamefont {T.~J.}\ \bibnamefont {Weiler}},\ }\href@noop {}
  {\bibfield  {journal} {\bibinfo  {journal} {Phys. Rev. Lett.}\ }\textbf
  {\bibinfo {volume} {90}},\ \bibinfo {pages} {181301} (\bibinfo {year}
  {2003})},\ \Eprint {http://arxiv.org/abs/hep-ph/0211305}
  {arXiv:hep-ph/0211305} \BibitemShut {NoStop}%
\bibitem [{\citenamefont {Meloni}\ and\ \citenamefont
  {Ohlsson}(2007)}]{Meloni:2006gv}%
  \BibitemOpen
  \bibfield  {author} {\bibinfo {author} {\bibfnamefont {D.}~\bibnamefont
  {Meloni}}\ and\ \bibinfo {author} {\bibfnamefont {T.}~\bibnamefont
  {Ohlsson}},\ }\href {\doibase 10.1103/PhysRevD.75.125017} {\bibfield
  {journal} {\bibinfo  {journal} {Phys. Rev. D}\ }\textbf {\bibinfo {volume}
  {75}},\ \bibinfo {pages} {125017} (\bibinfo {year} {2007})},\ \Eprint
  {http://arxiv.org/abs/hep-ph/0612279} {arXiv:hep-ph/0612279} \BibitemShut
  {NoStop}%
\bibitem [{\citenamefont {Baerwald}\ \emph {et~al.}(2012)\citenamefont
  {Baerwald}, \citenamefont {Bustamante},\ and\ \citenamefont
  {Winter}}]{Baerwald:2012kc}%
  \BibitemOpen
  \bibfield  {author} {\bibinfo {author} {\bibfnamefont {P.}~\bibnamefont
  {Baerwald}}, \bibinfo {author} {\bibfnamefont {M.}~\bibnamefont
  {Bustamante}}, \ and\ \bibinfo {author} {\bibfnamefont {W.}~\bibnamefont
  {Winter}},\ }\href@noop {} {\bibfield  {journal} {\bibinfo  {journal} {JCAP}\
  }\textbf {\bibinfo {volume} {10}},\ \bibinfo {pages} {020} (\bibinfo {year}
  {2012})},\ \Eprint {http://arxiv.org/abs/1208.4600} {arXiv:1208.4600
  [astro-ph.CO]} \BibitemShut {NoStop}%
\bibitem [{\citenamefont {Pagliaroli}\ \emph {et~al.}(2015)\citenamefont
  {Pagliaroli}, \citenamefont {Palladino}, \citenamefont {Villante},\ and\
  \citenamefont {Vissani}}]{Pagliaroli:2015rca}%
  \BibitemOpen
  \bibfield  {author} {\bibinfo {author} {\bibfnamefont {G.}~\bibnamefont
  {Pagliaroli}}, \bibinfo {author} {\bibfnamefont {A.}~\bibnamefont
  {Palladino}}, \bibinfo {author} {\bibfnamefont {F.~L.}\ \bibnamefont
  {Villante}}, \ and\ \bibinfo {author} {\bibfnamefont {F.}~\bibnamefont
  {Vissani}},\ }\href@noop {} {\bibfield  {journal} {\bibinfo  {journal} {Phys.
  Rev. D}\ }\textbf {\bibinfo {volume} {92}},\ \bibinfo {pages} {113008}
  (\bibinfo {year} {2015})},\ \Eprint {http://arxiv.org/abs/1506.02624}
  {arXiv:1506.02624 [hep-ph]} \BibitemShut {NoStop}%
\bibitem [{\citenamefont {Bustamante}\ \emph {et~al.}(2017)\citenamefont
  {Bustamante}, \citenamefont {Beacom},\ and\ \citenamefont
  {Murase}}]{Bustamante:2016ciw}%
  \BibitemOpen
  \bibfield  {author} {\bibinfo {author} {\bibfnamefont {M.}~\bibnamefont
  {Bustamante}}, \bibinfo {author} {\bibfnamefont {J.~F.}\ \bibnamefont
  {Beacom}}, \ and\ \bibinfo {author} {\bibfnamefont {K.}~\bibnamefont
  {Murase}},\ }\href@noop {} {\bibfield  {journal} {\bibinfo  {journal} {Phys.
  Rev. D}\ }\textbf {\bibinfo {volume} {95}},\ \bibinfo {pages} {063013}
  (\bibinfo {year} {2017})},\ \Eprint {http://arxiv.org/abs/1610.02096}
  {arXiv:1610.02096 [astro-ph.HE]} \BibitemShut {NoStop}%
\bibitem [{\citenamefont {Denton}\ and\ \citenamefont
  {Tamborra}(2018)}]{Denton:2018aml}%
  \BibitemOpen
  \bibfield  {author} {\bibinfo {author} {\bibfnamefont {P.~B.}\ \bibnamefont
  {Denton}}\ and\ \bibinfo {author} {\bibfnamefont {I.}~\bibnamefont
  {Tamborra}},\ }\href@noop {} {\bibfield  {journal} {\bibinfo  {journal}
  {Phys. Rev. Lett.}\ }\textbf {\bibinfo {volume} {121}},\ \bibinfo {pages}
  {121802} (\bibinfo {year} {2018})},\ \Eprint
  {http://arxiv.org/abs/1805.05950} {arXiv:1805.05950 [hep-ph]} \BibitemShut
  {NoStop}%
\bibitem [{\citenamefont {Huang}\ and\ \citenamefont
  {Zhou}(2024)}]{Huang:2024tbo}%
  \BibitemOpen
  \bibfield  {author} {\bibinfo {author} {\bibfnamefont {J.}~\bibnamefont
  {Huang}}\ and\ \bibinfo {author} {\bibfnamefont {S.}~\bibnamefont {Zhou}},\
  }\href@noop {} {\  (\bibinfo {year} {2024})},\ \Eprint
  {http://arxiv.org/abs/2407.04932} {arXiv:2407.04932 [hep-ph]} \BibitemShut
  {NoStop}%
\bibitem [{\citenamefont {Valera}\ \emph {et~al.}(2024)\citenamefont {Valera},
  \citenamefont {Fiorillo}, \citenamefont {Esteban},\ and\ \citenamefont
  {Bustamante}}]{Valera:2024buc}%
  \BibitemOpen
  \bibfield  {author} {\bibinfo {author} {\bibfnamefont {V.~B.}\ \bibnamefont
  {Valera}}, \bibinfo {author} {\bibfnamefont {D.~F.~G.}\ \bibnamefont
  {Fiorillo}}, \bibinfo {author} {\bibfnamefont {I.}~\bibnamefont {Esteban}}, \
  and\ \bibinfo {author} {\bibfnamefont {M.}~\bibnamefont {Bustamante}},\
  }\href {\doibase 10.1103/PhysRevD.110.043004} {\bibfield  {journal} {\bibinfo
   {journal} {Phys. Rev. D}\ }\textbf {\bibinfo {volume} {110}},\ \bibinfo
  {pages} {043004} (\bibinfo {year} {2024})},\ \Eprint
  {http://arxiv.org/abs/2405.14826} {arXiv:2405.14826 [astro-ph.HE]}
  \BibitemShut {NoStop}%
\bibitem [{\citenamefont {Kobayashi}\ and\ \citenamefont
  {Lim}(2001)}]{Kobayashi:2000md}%
  \BibitemOpen
  \bibfield  {author} {\bibinfo {author} {\bibfnamefont {M.}~\bibnamefont
  {Kobayashi}}\ and\ \bibinfo {author} {\bibfnamefont {C.~S.}\ \bibnamefont
  {Lim}},\ }\href@noop {} {\bibfield  {journal} {\bibinfo  {journal} {Phys.
  Rev. D}\ }\textbf {\bibinfo {volume} {64}},\ \bibinfo {pages} {013003}
  (\bibinfo {year} {2001})},\ \Eprint {http://arxiv.org/abs/hep-ph/0012266}
  {arXiv:hep-ph/0012266} \BibitemShut {NoStop}%
\bibitem [{\citenamefont {Beacom}\ \emph {et~al.}(2004)\citenamefont {Beacom},
  \citenamefont {Bell}, \citenamefont {Hooper}, \citenamefont {Learned},
  \citenamefont {Pakvasa},\ and\ \citenamefont {Weiler}}]{Beacom:2003eu}%
  \BibitemOpen
  \bibfield  {author} {\bibinfo {author} {\bibfnamefont {J.~F.}\ \bibnamefont
  {Beacom}}, \bibinfo {author} {\bibfnamefont {N.~F.}\ \bibnamefont {Bell}},
  \bibinfo {author} {\bibfnamefont {D.}~\bibnamefont {Hooper}}, \bibinfo
  {author} {\bibfnamefont {J.~G.}\ \bibnamefont {Learned}}, \bibinfo {author}
  {\bibfnamefont {S.}~\bibnamefont {Pakvasa}}, \ and\ \bibinfo {author}
  {\bibfnamefont {T.~J.}\ \bibnamefont {Weiler}},\ }\href@noop {} {\bibfield
  {journal} {\bibinfo  {journal} {Phys. Rev. Lett.}\ }\textbf {\bibinfo
  {volume} {92}},\ \bibinfo {pages} {011101} (\bibinfo {year} {2004})},\
  \Eprint {http://arxiv.org/abs/hep-ph/0307151} {arXiv:hep-ph/0307151}
  \BibitemShut {NoStop}%
\bibitem [{\citenamefont {Carloni}\ \emph {et~al.}(2024)\citenamefont
  {Carloni}, \citenamefont {Mart\'\i{}nez-Soler}, \citenamefont {Arguelles},
  \citenamefont {Babu},\ and\ \citenamefont {Dev}}]{Carloni:2022cqz}%
  \BibitemOpen
  \bibfield  {author} {\bibinfo {author} {\bibfnamefont {K.}~\bibnamefont
  {Carloni}}, \bibinfo {author} {\bibfnamefont {I.}~\bibnamefont
  {Mart\'\i{}nez-Soler}}, \bibinfo {author} {\bibfnamefont {C.~A.}\
  \bibnamefont {Arguelles}}, \bibinfo {author} {\bibfnamefont {K.~S.}\
  \bibnamefont {Babu}}, \ and\ \bibinfo {author} {\bibfnamefont {P.~S.~B.}\
  \bibnamefont {Dev}},\ }\href {\doibase 10.1103/PhysRevD.109.L051702}
  {\bibfield  {journal} {\bibinfo  {journal} {Phys. Rev. D}\ }\textbf {\bibinfo
  {volume} {109}},\ \bibinfo {pages} {L051702} (\bibinfo {year} {2024})},\
  \Eprint {http://arxiv.org/abs/2212.00737} {arXiv:2212.00737 [astro-ph.HE]}
  \BibitemShut {NoStop}%
\bibitem [{\citenamefont {Rink}\ and\ \citenamefont
  {Sen}(2024)}]{Rink:2022nvw}%
  \BibitemOpen
  \bibfield  {author} {\bibinfo {author} {\bibfnamefont {T.}~\bibnamefont
  {Rink}}\ and\ \bibinfo {author} {\bibfnamefont {M.}~\bibnamefont {Sen}},\
  }\href {\doibase 10.1016/j.physletb.2024.138558} {\bibfield  {journal}
  {\bibinfo  {journal} {Phys. Lett. B}\ }\textbf {\bibinfo {volume} {851}},\
  \bibinfo {pages} {138558} (\bibinfo {year} {2024})},\ \Eprint
  {http://arxiv.org/abs/2211.16520} {arXiv:2211.16520 [hep-ph]} \BibitemShut
  {NoStop}%
\bibitem [{\citenamefont {Franklin}\ \emph {et~al.}(2023)\citenamefont
  {Franklin}, \citenamefont {Perez-Gonzalez},\ and\ \citenamefont
  {Turner}}]{Franklin:2023diy}%
  \BibitemOpen
  \bibfield  {author} {\bibinfo {author} {\bibfnamefont {J.}~\bibnamefont
  {Franklin}}, \bibinfo {author} {\bibfnamefont {Y.~F.}\ \bibnamefont
  {Perez-Gonzalez}}, \ and\ \bibinfo {author} {\bibfnamefont {J.}~\bibnamefont
  {Turner}},\ }\href {\doibase 10.1103/PhysRevD.108.035010} {\bibfield
  {journal} {\bibinfo  {journal} {Phys. Rev. D}\ }\textbf {\bibinfo {volume}
  {108}},\ \bibinfo {pages} {035010} (\bibinfo {year} {2023})},\ \Eprint
  {http://arxiv.org/abs/2304.05418} {arXiv:2304.05418 [hep-ph]} \BibitemShut
  {NoStop}%
\bibitem [{\citenamefont {Antusch}\ and\ \citenamefont
  {Fischer}(2014)}]{Antusch:2014woa}%
  \BibitemOpen
  \bibfield  {author} {\bibinfo {author} {\bibfnamefont {S.}~\bibnamefont
  {Antusch}}\ and\ \bibinfo {author} {\bibfnamefont {O.}~\bibnamefont
  {Fischer}},\ }\href {\doibase 10.1007/JHEP10(2014)094} {\bibfield  {journal}
  {\bibinfo  {journal} {JHEP}\ }\textbf {\bibinfo {volume} {10}},\ \bibinfo
  {pages} {094} (\bibinfo {year} {2014})},\ \Eprint
  {http://arxiv.org/abs/1407.6607} {arXiv:1407.6607 [hep-ph]} \BibitemShut
  {NoStop}%
\bibitem [{\citenamefont {Blennow}\ \emph {et~al.}(2023)\citenamefont
  {Blennow}, \citenamefont {Fern\'andez-Mart\'\i{}nez}, \citenamefont
  {Hern\'andez-Garc\'\i{}a}, \citenamefont {L\'opez-Pav\'on}, \citenamefont
  {Marcano},\ and\ \citenamefont {Naredo-Tuero}}]{Blennow:2023mqx}%
  \BibitemOpen
  \bibfield  {author} {\bibinfo {author} {\bibfnamefont {M.}~\bibnamefont
  {Blennow}}, \bibinfo {author} {\bibfnamefont {E.}~\bibnamefont
  {Fern\'andez-Mart\'\i{}nez}}, \bibinfo {author} {\bibfnamefont
  {J.}~\bibnamefont {Hern\'andez-Garc\'\i{}a}}, \bibinfo {author}
  {\bibfnamefont {J.}~\bibnamefont {L\'opez-Pav\'on}}, \bibinfo {author}
  {\bibfnamefont {X.}~\bibnamefont {Marcano}}, \ and\ \bibinfo {author}
  {\bibfnamefont {D.}~\bibnamefont {Naredo-Tuero}},\ }\href {\doibase
  10.1007/JHEP08(2023)030} {\bibfield  {journal} {\bibinfo  {journal} {JHEP}\
  }\textbf {\bibinfo {volume} {08}},\ \bibinfo {pages} {030} (\bibinfo {year}
  {2023})},\ \Eprint {http://arxiv.org/abs/2306.01040} {arXiv:2306.01040
  [hep-ph]} \BibitemShut {NoStop}%
\bibitem [{\citenamefont {Aloni}\ and\ \citenamefont
  {Dery}(2024)}]{Aloni:2022ebm}%
  \BibitemOpen
  \bibfield  {author} {\bibinfo {author} {\bibfnamefont {D.}~\bibnamefont
  {Aloni}}\ and\ \bibinfo {author} {\bibfnamefont {A.}~\bibnamefont {Dery}},\
  }\href {\doibase 10.1103/PhysRevD.109.055006} {\bibfield  {journal} {\bibinfo
   {journal} {Phys. Rev. D}\ }\textbf {\bibinfo {volume} {109}},\ \bibinfo
  {pages} {055006} (\bibinfo {year} {2024})},\ \Eprint
  {http://arxiv.org/abs/2211.09638} {arXiv:2211.09638 [hep-ph]} \BibitemShut
  {NoStop}%
\bibitem [{\citenamefont {Huitu}\ \emph {et~al.}(2016)\citenamefont {Huitu},
  \citenamefont {K\"arkk\"ainen}, \citenamefont {Maalampi},\ and\ \citenamefont
  {Vihonen}}]{Huitu:2016bmb}%
  \BibitemOpen
  \bibfield  {author} {\bibinfo {author} {\bibfnamefont {K.}~\bibnamefont
  {Huitu}}, \bibinfo {author} {\bibfnamefont {T.~J.}\ \bibnamefont
  {K\"arkk\"ainen}}, \bibinfo {author} {\bibfnamefont {J.}~\bibnamefont
  {Maalampi}}, \ and\ \bibinfo {author} {\bibfnamefont {S.}~\bibnamefont
  {Vihonen}},\ }\href {\doibase 10.1103/PhysRevD.93.053016} {\bibfield
  {journal} {\bibinfo  {journal} {Phys. Rev. D}\ }\textbf {\bibinfo {volume}
  {93}},\ \bibinfo {pages} {053016} (\bibinfo {year} {2016})},\ \Eprint
  {http://arxiv.org/abs/1601.07730} {arXiv:1601.07730 [hep-ph]} \BibitemShut
  {NoStop}%
\bibitem [{\citenamefont {Adamson}\ \emph {et~al.}(2017)\citenamefont {Adamson}
  \emph {et~al.}}]{MINOS:2016sbv}%
  \BibitemOpen
  \bibfield  {author} {\bibinfo {author} {\bibfnamefont {P.}~\bibnamefont
  {Adamson}} \emph {et~al.} (\bibinfo {collaboration} {MINOS}),\ }\href
  {\doibase 10.1103/PhysRevD.95.012005} {\bibfield  {journal} {\bibinfo
  {journal} {Phys. Rev. D}\ }\textbf {\bibinfo {volume} {95}},\ \bibinfo
  {pages} {012005} (\bibinfo {year} {2017})},\ \Eprint
  {http://arxiv.org/abs/1605.06169} {arXiv:1605.06169 [hep-ex]} \BibitemShut
  {NoStop}%
\bibitem [{\citenamefont {Esteban}\ \emph {et~al.}(2018)\citenamefont
  {Esteban}, \citenamefont {Gonzalez-Garcia}, \citenamefont {Maltoni},
  \citenamefont {Martinez-Soler},\ and\ \citenamefont
  {Salvado}}]{Esteban:2018ppq}%
  \BibitemOpen
  \bibfield  {author} {\bibinfo {author} {\bibfnamefont {I.}~\bibnamefont
  {Esteban}}, \bibinfo {author} {\bibfnamefont {M.~C.}\ \bibnamefont
  {Gonzalez-Garcia}}, \bibinfo {author} {\bibfnamefont {M.}~\bibnamefont
  {Maltoni}}, \bibinfo {author} {\bibfnamefont {I.}~\bibnamefont
  {Martinez-Soler}}, \ and\ \bibinfo {author} {\bibfnamefont {J.}~\bibnamefont
  {Salvado}},\ }\href {\doibase 10.1007/JHEP08(2018)180} {\bibfield  {journal}
  {\bibinfo  {journal} {JHEP}\ }\textbf {\bibinfo {volume} {08}},\ \bibinfo
  {pages} {180} (\bibinfo {year} {2018})},\ \bibinfo {note} {[Addendum: JHEP
  12, 152 (2020)]},\ \Eprint {http://arxiv.org/abs/1805.04530}
  {arXiv:1805.04530 [hep-ph]} \BibitemShut {NoStop}%
\bibitem [{\citenamefont {Agarwalla}\ \emph {et~al.}(2020)\citenamefont
  {Agarwalla} \emph {et~al.}}]{Borexino:2019mhy}%
  \BibitemOpen
  \bibfield  {author} {\bibinfo {author} {\bibfnamefont {S.~K.}\ \bibnamefont
  {Agarwalla}} \emph {et~al.} (\bibinfo {collaboration} {Borexino}),\ }\href
  {\doibase 10.1007/JHEP02(2020)038} {\bibfield  {journal} {\bibinfo  {journal}
  {JHEP}\ }\textbf {\bibinfo {volume} {02}},\ \bibinfo {pages} {038} (\bibinfo
  {year} {2020})},\ \Eprint {http://arxiv.org/abs/1905.03512} {arXiv:1905.03512
  [hep-ph]} \BibitemShut {NoStop}%
\bibitem [{\citenamefont {Masud}\ \emph {et~al.}(2021)\citenamefont {Masud},
  \citenamefont {Mehta}, \citenamefont {Ternes},\ and\ \citenamefont
  {Tortola}}]{Masud:2021ves}%
  \BibitemOpen
  \bibfield  {author} {\bibinfo {author} {\bibfnamefont {M.}~\bibnamefont
  {Masud}}, \bibinfo {author} {\bibfnamefont {P.}~\bibnamefont {Mehta}},
  \bibinfo {author} {\bibfnamefont {C.~A.}\ \bibnamefont {Ternes}}, \ and\
  \bibinfo {author} {\bibfnamefont {M.}~\bibnamefont {Tortola}},\ }\href
  {\doibase 10.1007/JHEP05(2021)171} {\bibfield  {journal} {\bibinfo  {journal}
  {JHEP}\ }\textbf {\bibinfo {volume} {05}},\ \bibinfo {pages} {171} (\bibinfo
  {year} {2021})},\ \Eprint {http://arxiv.org/abs/2103.11143} {arXiv:2103.11143
  [hep-ph]} \BibitemShut {NoStop}%
\bibitem [{\citenamefont {Abbasi}\ \emph
  {et~al.}(2021{\natexlab{a}})\citenamefont {Abbasi} \emph
  {et~al.}}]{IceCubeCollaboration:2021euf}%
  \BibitemOpen
  \bibfield  {author} {\bibinfo {author} {\bibfnamefont {R.}~\bibnamefont
  {Abbasi}} \emph {et~al.} (\bibinfo {collaboration} {(IceCube Collaboration)*,
  IceCube}),\ }\href {\doibase 10.1103/PhysRevD.104.072006} {\bibfield
  {journal} {\bibinfo  {journal} {Phys. Rev. D}\ }\textbf {\bibinfo {volume}
  {104}},\ \bibinfo {pages} {072006} (\bibinfo {year} {2021}{\natexlab{a}})},\
  \Eprint {http://arxiv.org/abs/2106.07755} {arXiv:2106.07755 [hep-ex]}
  \BibitemShut {NoStop}%
\bibitem [{\citenamefont {Bakhti}\ and\ \citenamefont
  {Rajaee}(2021)}]{Bakhti:2020fde}%
  \BibitemOpen
  \bibfield  {author} {\bibinfo {author} {\bibfnamefont {P.}~\bibnamefont
  {Bakhti}}\ and\ \bibinfo {author} {\bibfnamefont {M.}~\bibnamefont
  {Rajaee}},\ }\href {\doibase 10.1103/PhysRevD.103.075003} {\bibfield
  {journal} {\bibinfo  {journal} {Phys. Rev. D}\ }\textbf {\bibinfo {volume}
  {103}},\ \bibinfo {pages} {075003} (\bibinfo {year} {2021})},\ \Eprint
  {http://arxiv.org/abs/2010.12849} {arXiv:2010.12849 [hep-ph]} \BibitemShut
  {NoStop}%
\bibitem [{\citenamefont {Brahma}\ and\ \citenamefont
  {Giri}(2022)}]{Brahma:2022xld}%
  \BibitemOpen
  \bibfield  {author} {\bibinfo {author} {\bibfnamefont {B.}~\bibnamefont
  {Brahma}}\ and\ \bibinfo {author} {\bibfnamefont {A.}~\bibnamefont {Giri}},\
  }\href {\doibase 10.1140/epjc/s10052-022-11134-x} {\bibfield  {journal}
  {\bibinfo  {journal} {Eur. Phys. J. C}\ }\textbf {\bibinfo {volume} {82}},\
  \bibinfo {pages} {1145} (\bibinfo {year} {2022})}\BibitemShut {NoStop}%
\bibitem [{\citenamefont {Lazo~Pedrajas}(2024)}]{LazoPedrajas:2024qlf}%
  \BibitemOpen
  \bibfield  {author} {\bibinfo {author} {\bibfnamefont {A.}~\bibnamefont
  {Lazo~Pedrajas}} (\bibinfo {collaboration} {KM3NeT}),\ }\href {\doibase
  10.22323/1.441.0193} {\bibfield  {journal} {\bibinfo  {journal} {PoS}\
  }\textbf {\bibinfo {volume} {TAUP2023}},\ \bibinfo {pages} {193} (\bibinfo
  {year} {2024})}\BibitemShut {NoStop}%
\bibitem [{\citenamefont {Fornengo}\ \emph {et~al.}(2002)\citenamefont
  {Fornengo}, \citenamefont {Maltoni}, \citenamefont {Tomas},\ and\
  \citenamefont {Valle}}]{Fornengo:2001pm}%
  \BibitemOpen
  \bibfield  {author} {\bibinfo {author} {\bibfnamefont {N.}~\bibnamefont
  {Fornengo}}, \bibinfo {author} {\bibfnamefont {M.}~\bibnamefont {Maltoni}},
  \bibinfo {author} {\bibfnamefont {R.}~\bibnamefont {Tomas}}, \ and\ \bibinfo
  {author} {\bibfnamefont {J.~W.~F.}\ \bibnamefont {Valle}},\ }\href {\doibase
  10.1103/PhysRevD.65.013010} {\bibfield  {journal} {\bibinfo  {journal} {Phys.
  Rev. D}\ }\textbf {\bibinfo {volume} {65}},\ \bibinfo {pages} {013010}
  (\bibinfo {year} {2002})},\ \Eprint {http://arxiv.org/abs/hep-ph/0108043}
  {arXiv:hep-ph/0108043} \BibitemShut {NoStop}%
\bibitem [{\citenamefont {Liao}\ \emph {et~al.}(2024)\citenamefont {Liao},
  \citenamefont {Marfatia},\ and\ \citenamefont {Zhang}}]{Liao:2024qoe}%
  \BibitemOpen
  \bibfield  {author} {\bibinfo {author} {\bibfnamefont {J.}~\bibnamefont
  {Liao}}, \bibinfo {author} {\bibfnamefont {D.}~\bibnamefont {Marfatia}}, \
  and\ \bibinfo {author} {\bibfnamefont {J.}~\bibnamefont {Zhang}},\
  }\href@noop {} {\  (\bibinfo {year} {2024})},\ \Eprint
  {http://arxiv.org/abs/2408.06255} {arXiv:2408.06255 [hep-ph]} \BibitemShut
  {NoStop}%
\bibitem [{\citenamefont {Abbasi}\ \emph {et~al.}(2023)\citenamefont {Abbasi}
  \emph {et~al.}}]{IceCube:2023sov}%
  \BibitemOpen
  \bibfield  {author} {\bibinfo {author} {\bibfnamefont {R.}~\bibnamefont
  {Abbasi}} \emph {et~al.} (\bibinfo {collaboration} {IceCube}),\ }\href
  {\doibase 10.22323/1.444.1030} {\bibfield  {journal} {\bibinfo  {journal}
  {PoS}\ }\textbf {\bibinfo {volume} {ICRC2023}},\ \bibinfo {pages} {1030}
  (\bibinfo {year} {2023})},\ \Eprint {http://arxiv.org/abs/2307.13878}
  {arXiv:2307.13878 [astro-ph.HE]} \BibitemShut {NoStop}%
\bibitem [{\citenamefont {He}\ \emph {et~al.}(1991)\citenamefont {He},
  \citenamefont {Joshi}, \citenamefont {Lew},\ and\ \citenamefont
  {Volkas}}]{He:1991qd}%
  \BibitemOpen
  \bibfield  {author} {\bibinfo {author} {\bibfnamefont {X.-G.}\ \bibnamefont
  {He}}, \bibinfo {author} {\bibfnamefont {G.~C.}\ \bibnamefont {Joshi}},
  \bibinfo {author} {\bibfnamefont {H.}~\bibnamefont {Lew}}, \ and\ \bibinfo
  {author} {\bibfnamefont {R.~R.}\ \bibnamefont {Volkas}},\ }\href {\doibase
  10.1103/PhysRevD.44.2118} {\bibfield  {journal} {\bibinfo  {journal} {Phys.
  Rev. D}\ }\textbf {\bibinfo {volume} {44}},\ \bibinfo {pages} {2118}
  (\bibinfo {year} {1991})}\BibitemShut {NoStop}%
\bibitem [{\citenamefont {Bustamante}\ and\ \citenamefont
  {Agarwalla}(2019)}]{Bustamante:2018mzu}%
  \BibitemOpen
  \bibfield  {author} {\bibinfo {author} {\bibfnamefont {M.}~\bibnamefont
  {Bustamante}}\ and\ \bibinfo {author} {\bibfnamefont {S.~K.}\ \bibnamefont
  {Agarwalla}},\ }\href {\doibase 10.1103/PhysRevLett.122.061103} {\bibfield
  {journal} {\bibinfo  {journal} {Phys. Rev. Lett.}\ }\textbf {\bibinfo
  {volume} {122}},\ \bibinfo {pages} {061103} (\bibinfo {year} {2019})},\
  \Eprint {http://arxiv.org/abs/1808.02042} {arXiv:1808.02042 [astro-ph.HE]}
  \BibitemShut {NoStop}%
\bibitem [{\citenamefont {Coloma}\ \emph {et~al.}(2021)\citenamefont {Coloma},
  \citenamefont {Gonzalez-Garcia},\ and\ \citenamefont
  {Maltoni}}]{Coloma:2020gfv}%
  \BibitemOpen
  \bibfield  {author} {\bibinfo {author} {\bibfnamefont {P.}~\bibnamefont
  {Coloma}}, \bibinfo {author} {\bibfnamefont {M.~C.}\ \bibnamefont
  {Gonzalez-Garcia}}, \ and\ \bibinfo {author} {\bibfnamefont {M.}~\bibnamefont
  {Maltoni}},\ }\href {\doibase 10.1007/JHEP01(2021)114} {\bibfield  {journal}
  {\bibinfo  {journal} {JHEP}\ }\textbf {\bibinfo {volume} {01}},\ \bibinfo
  {pages} {114} (\bibinfo {year} {2021})},\ \bibinfo {note} {[Erratum: JHEP 11,
  115 (2022)]},\ \Eprint {http://arxiv.org/abs/2009.14220} {arXiv:2009.14220
  [hep-ph]} \BibitemShut {NoStop}%
\bibitem [{\citenamefont {Agarwalla}\ \emph {et~al.}(2023)\citenamefont
  {Agarwalla}, \citenamefont {Bustamante}, \citenamefont {Das},\ and\
  \citenamefont {Narang}}]{Agarwalla:2023sng}%
  \BibitemOpen
  \bibfield  {author} {\bibinfo {author} {\bibfnamefont {S.~K.}\ \bibnamefont
  {Agarwalla}}, \bibinfo {author} {\bibfnamefont {M.}~\bibnamefont
  {Bustamante}}, \bibinfo {author} {\bibfnamefont {S.}~\bibnamefont {Das}}, \
  and\ \bibinfo {author} {\bibfnamefont {A.}~\bibnamefont {Narang}},\ }\href
  {\doibase 10.1007/JHEP08(2023)113} {\bibfield  {journal} {\bibinfo  {journal}
  {JHEP}\ }\textbf {\bibinfo {volume} {08}},\ \bibinfo {pages} {113} (\bibinfo
  {year} {2023})},\ \Eprint {http://arxiv.org/abs/2305.03675} {arXiv:2305.03675
  [hep-ph]} \BibitemShut {NoStop}%
\bibitem [{\citenamefont {Agarwalla}\ \emph {et~al.}(2024)\citenamefont
  {Agarwalla}, \citenamefont {Bustamante}, \citenamefont {Singh},\ and\
  \citenamefont {Swain}}]{Agarwalla:2024ylc}%
  \BibitemOpen
  \bibfield  {author} {\bibinfo {author} {\bibfnamefont {S.~K.}\ \bibnamefont
  {Agarwalla}}, \bibinfo {author} {\bibfnamefont {M.}~\bibnamefont
  {Bustamante}}, \bibinfo {author} {\bibfnamefont {M.}~\bibnamefont {Singh}}, \
  and\ \bibinfo {author} {\bibfnamefont {P.}~\bibnamefont {Swain}},\
  }\href@noop {} {\  (\bibinfo {year} {2024})},\ \Eprint
  {http://arxiv.org/abs/2404.02775} {arXiv:2404.02775 [hep-ph]} \BibitemShut
  {NoStop}%
\bibitem [{\citenamefont {Glashow}(1960)}]{Glashow:1960zz}%
  \BibitemOpen
  \bibfield  {author} {\bibinfo {author} {\bibfnamefont {S.~L.}\ \bibnamefont
  {Glashow}},\ }\href@noop {} {\bibfield  {journal} {\bibinfo  {journal} {Phys.
  Rev.}\ }\textbf {\bibinfo {volume} {118}},\ \bibinfo {pages} {316} (\bibinfo
  {year} {1960})}\BibitemShut {NoStop}%
\bibitem [{\citenamefont {Barger}\ \emph {et~al.}(2014)\citenamefont {Barger},
  \citenamefont {Fu}, \citenamefont {Learned}, \citenamefont {Marfatia},
  \citenamefont {Pakvasa},\ and\ \citenamefont {Weiler}}]{Barger:2014iua}%
  \BibitemOpen
  \bibfield  {author} {\bibinfo {author} {\bibfnamefont {V.}~\bibnamefont
  {Barger}}, \bibinfo {author} {\bibfnamefont {L.}~\bibnamefont {Fu}}, \bibinfo
  {author} {\bibfnamefont {J.~G.}\ \bibnamefont {Learned}}, \bibinfo {author}
  {\bibfnamefont {D.}~\bibnamefont {Marfatia}}, \bibinfo {author}
  {\bibfnamefont {S.}~\bibnamefont {Pakvasa}}, \ and\ \bibinfo {author}
  {\bibfnamefont {T.~J.}\ \bibnamefont {Weiler}},\ }\href {\doibase
  10.1103/PhysRevD.90.121301} {\bibfield  {journal} {\bibinfo  {journal} {Phys.
  Rev. D}\ }\textbf {\bibinfo {volume} {90}},\ \bibinfo {pages} {121301}
  (\bibinfo {year} {2014})},\ \Eprint {http://arxiv.org/abs/1407.3255}
  {arXiv:1407.3255 [astro-ph.HE]} \BibitemShut {NoStop}%
\bibitem [{\citenamefont {Huang}\ and\ \citenamefont
  {Liu}(2020)}]{Huang:2019hgs}%
  \BibitemOpen
  \bibfield  {author} {\bibinfo {author} {\bibfnamefont {G.-y.}\ \bibnamefont
  {Huang}}\ and\ \bibinfo {author} {\bibfnamefont {Q.}~\bibnamefont {Liu}},\
  }\href@noop {} {\bibfield  {journal} {\bibinfo  {journal} {JCAP}\ }\textbf
  {\bibinfo {volume} {03}},\ \bibinfo {pages} {005} (\bibinfo {year} {2020})},\
  \Eprint {http://arxiv.org/abs/1912.02976} {arXiv:1912.02976 [hep-ph]}
  \BibitemShut {NoStop}%
\bibitem [{\citenamefont {Xu}\ and\ \citenamefont {Rong}(2023)}]{Xu:2022svm}%
  \BibitemOpen
  \bibfield  {author} {\bibinfo {author} {\bibfnamefont {D.-H.}\ \bibnamefont
  {Xu}}\ and\ \bibinfo {author} {\bibfnamefont {S.-J.}\ \bibnamefont {Rong}},\
  }\href {\doibase 10.1142/S0217732323500293} {\bibfield  {journal} {\bibinfo
  {journal} {Mod. Phys. Lett. A}\ }\textbf {\bibinfo {volume} {38}},\ \bibinfo
  {pages} {2350029} (\bibinfo {year} {2023})},\ \Eprint
  {http://arxiv.org/abs/2211.05478} {arXiv:2211.05478 [hep-ph]} \BibitemShut
  {NoStop}%
\bibitem [{\citenamefont {Abbasi}\ \emph {et~al.}(2024)\citenamefont {Abbasi}
  \emph {et~al.}}]{IceCube:2024nhk}%
  \BibitemOpen
  \bibfield  {author} {\bibinfo {author} {\bibfnamefont {R.}~\bibnamefont
  {Abbasi}} \emph {et~al.} (\bibinfo {collaboration} {IceCube}),\ }\href
  {\doibase 10.1103/PhysRevLett.132.151001} {\bibfield  {journal} {\bibinfo
  {journal} {Phys. Rev. Lett.}\ }\textbf {\bibinfo {volume} {132}},\ \bibinfo
  {pages} {151001} (\bibinfo {year} {2024})},\ \Eprint
  {http://arxiv.org/abs/2403.02516} {arXiv:2403.02516 [astro-ph.HE]}
  \BibitemShut {NoStop}%
\bibitem [{\citenamefont {Agostini}\ \emph {et~al.}(2020)\citenamefont
  {Agostini} \emph {et~al.}}]{P-ONE:2020ljt}%
  \BibitemOpen
  \bibfield  {author} {\bibinfo {author} {\bibfnamefont {M.}~\bibnamefont
  {Agostini}} \emph {et~al.} (\bibinfo {collaboration} {P-ONE}),\ }\href
  {\doibase 10.1038/s41550-020-1182-4} {\bibfield  {journal} {\bibinfo
  {journal} {Nature Astron.}\ }\textbf {\bibinfo {volume} {4}},\ \bibinfo
  {pages} {913} (\bibinfo {year} {2020})},\ \Eprint
  {http://arxiv.org/abs/2005.09493} {arXiv:2005.09493 [astro-ph.HE]}
  \BibitemShut {NoStop}%
\bibitem [{\citenamefont {Aartsen}\ \emph {et~al.}(2021)\citenamefont {Aartsen}
  \emph {et~al.}}]{IceCube-Gen2:2020qha}%
  \BibitemOpen
  \bibfield  {author} {\bibinfo {author} {\bibfnamefont {M.~G.}\ \bibnamefont
  {Aartsen}} \emph {et~al.} (\bibinfo {collaboration} {IceCube-Gen2}),\ }\href
  {\doibase 10.1088/1361-6471/abbd48} {\bibfield  {journal} {\bibinfo
  {journal} {J. Phys. G}\ }\textbf {\bibinfo {volume} {48}},\ \bibinfo {pages}
  {060501} (\bibinfo {year} {2021})},\ \Eprint
  {http://arxiv.org/abs/2008.04323} {arXiv:2008.04323 [astro-ph.HE]}
  \BibitemShut {NoStop}%
\bibitem [{\citenamefont {Wolfenstein}(1978)}]{Wolfenstein:1977ue}%
  \BibitemOpen
  \bibfield  {author} {\bibinfo {author} {\bibfnamefont {L.}~\bibnamefont
  {Wolfenstein}},\ }\href {\doibase 10.1103/PhysRevD.17.2369} {\bibfield
  {journal} {\bibinfo  {journal} {Phys. Rev. D}\ }\textbf {\bibinfo {volume}
  {17}},\ \bibinfo {pages} {2369} (\bibinfo {year} {1978})}\BibitemShut
  {NoStop}%
\bibitem [{\citenamefont {Mikheyev}\ and\ \citenamefont
  {Smirnov}(1985)}]{Mikheyev:1985zog}%
  \BibitemOpen
  \bibfield  {author} {\bibinfo {author} {\bibfnamefont {S.~P.}\ \bibnamefont
  {Mikheyev}}\ and\ \bibinfo {author} {\bibfnamefont {A.~Y.}\ \bibnamefont
  {Smirnov}},\ }\href@noop {} {\bibfield  {journal} {\bibinfo  {journal} {Sov.
  J. Nucl. Phys.}\ }\textbf {\bibinfo {volume} {42}},\ \bibinfo {pages} {913}
  (\bibinfo {year} {1985})}\BibitemShut {NoStop}%
\bibitem [{\citenamefont {Halzen}\ and\ \citenamefont
  {Wille}(2016{\natexlab{a}})}]{Halzen:2016pwl}%
  \BibitemOpen
  \bibfield  {author} {\bibinfo {author} {\bibfnamefont {F.}~\bibnamefont
  {Halzen}}\ and\ \bibinfo {author} {\bibfnamefont {L.}~\bibnamefont {Wille}},\
  }\href@noop {} {\  (\bibinfo {year} {2016}{\natexlab{a}})},\ \Eprint
  {http://arxiv.org/abs/1601.03044} {arXiv:1601.03044 [hep-ph]} \BibitemShut
  {NoStop}%
\bibitem [{\citenamefont {Halzen}\ and\ \citenamefont
  {Wille}(2016{\natexlab{b}})}]{Halzen:2016thi}%
  \BibitemOpen
  \bibfield  {author} {\bibinfo {author} {\bibfnamefont {F.}~\bibnamefont
  {Halzen}}\ and\ \bibinfo {author} {\bibfnamefont {L.}~\bibnamefont {Wille}},\
  }\href {\doibase 10.1103/PhysRevD.94.014014} {\bibfield  {journal} {\bibinfo
  {journal} {Phys. Rev. D}\ }\textbf {\bibinfo {volume} {94}},\ \bibinfo
  {pages} {014014} (\bibinfo {year} {2016}{\natexlab{b}})},\ \Eprint
  {http://arxiv.org/abs/1605.01409} {arXiv:1605.01409 [hep-ph]} \BibitemShut
  {NoStop}%
\bibitem [{\citenamefont {Aartsen}\ \emph
  {et~al.}(2015{\natexlab{a}})\citenamefont {Aartsen} \emph
  {et~al.}}]{IceCube:2015rro}%
  \BibitemOpen
  \bibfield  {author} {\bibinfo {author} {\bibfnamefont {M.~G.}\ \bibnamefont
  {Aartsen}} \emph {et~al.} (\bibinfo {collaboration} {IceCube}),\ }\href
  {\doibase 10.1103/PhysRevLett.114.171102} {\bibfield  {journal} {\bibinfo
  {journal} {Phys. Rev. Lett.}\ }\textbf {\bibinfo {volume} {114}},\ \bibinfo
  {pages} {171102} (\bibinfo {year} {2015}{\natexlab{a}})},\ \Eprint
  {http://arxiv.org/abs/1502.03376} {arXiv:1502.03376 [astro-ph.HE]}
  \BibitemShut {NoStop}%
\bibitem [{\citenamefont {Aartsen}\ \emph
  {et~al.}(2015{\natexlab{b}})\citenamefont {Aartsen} \emph
  {et~al.}}]{IceCube:2015qii}%
  \BibitemOpen
  \bibfield  {author} {\bibinfo {author} {\bibfnamefont {M.~G.}\ \bibnamefont
  {Aartsen}} \emph {et~al.} (\bibinfo {collaboration} {IceCube}),\ }\href
  {\doibase 10.1103/PhysRevLett.115.081102} {\bibfield  {journal} {\bibinfo
  {journal} {Phys. Rev. Lett.}\ }\textbf {\bibinfo {volume} {115}},\ \bibinfo
  {pages} {081102} (\bibinfo {year} {2015}{\natexlab{b}})},\ \Eprint
  {http://arxiv.org/abs/1507.04005} {arXiv:1507.04005 [astro-ph.HE]}
  \BibitemShut {NoStop}%
\bibitem [{\citenamefont {Kopper}(2018)}]{Kopper:2017zzm}%
  \BibitemOpen
  \bibfield  {author} {\bibinfo {author} {\bibfnamefont {C.}~\bibnamefont
  {Kopper}} (\bibinfo {collaboration} {IceCube}),\ }\href {\doibase
  10.22323/1.301.0981} {\bibfield  {journal} {\bibinfo  {journal} {PoS}\
  }\textbf {\bibinfo {volume} {ICRC2017}},\ \bibinfo {pages} {981} (\bibinfo
  {year} {2018})}\BibitemShut {NoStop}%
\bibitem [{\citenamefont {Abbasi}\ \emph
  {et~al.}(2021{\natexlab{b}})\citenamefont {Abbasi} \emph
  {et~al.}}]{IceCube:2020wum}%
  \BibitemOpen
  \bibfield  {author} {\bibinfo {author} {\bibfnamefont {R.}~\bibnamefont
  {Abbasi}} \emph {et~al.} (\bibinfo {collaboration} {IceCube}),\ }\href
  {\doibase 10.1103/PhysRevD.104.022002} {\bibfield  {journal} {\bibinfo
  {journal} {Phys. Rev. D}\ }\textbf {\bibinfo {volume} {104}},\ \bibinfo
  {pages} {022002} (\bibinfo {year} {2021}{\natexlab{b}})},\ \Eprint
  {http://arxiv.org/abs/2011.03545} {arXiv:2011.03545 [astro-ph.HE]}
  \BibitemShut {NoStop}%
\bibitem [{\citenamefont {Esteban}\ \emph {et~al.}(2020)\citenamefont
  {Esteban}, \citenamefont {Gonzalez-Garcia}, \citenamefont {Maltoni},
  \citenamefont {Schwetz},\ and\ \citenamefont {Zhou}}]{Esteban:2020cvm}%
  \BibitemOpen
  \bibfield  {author} {\bibinfo {author} {\bibfnamefont {I.}~\bibnamefont
  {Esteban}}, \bibinfo {author} {\bibfnamefont {M.~C.}\ \bibnamefont
  {Gonzalez-Garcia}}, \bibinfo {author} {\bibfnamefont {M.}~\bibnamefont
  {Maltoni}}, \bibinfo {author} {\bibfnamefont {T.}~\bibnamefont {Schwetz}}, \
  and\ \bibinfo {author} {\bibfnamefont {A.}~\bibnamefont {Zhou}},\ }\href
  {\doibase 10.1007/JHEP09(2020)178} {\bibfield  {journal} {\bibinfo  {journal}
  {JHEP}\ }\textbf {\bibinfo {volume} {09}},\ \bibinfo {pages} {178} (\bibinfo
  {year} {2020})},\ \Eprint {http://arxiv.org/abs/2007.14792} {arXiv:2007.14792
  [hep-ph]} \BibitemShut {NoStop}%
\bibitem [{\citenamefont {Song}\ \emph {et~al.}(2021)\citenamefont {Song},
  \citenamefont {Li}, \citenamefont {Arg\"uelles}, \citenamefont {Bustamante},\
  and\ \citenamefont {Vincent}}]{Song:2020nfh}%
  \BibitemOpen
  \bibfield  {author} {\bibinfo {author} {\bibfnamefont {N.}~\bibnamefont
  {Song}}, \bibinfo {author} {\bibfnamefont {S.~W.}\ \bibnamefont {Li}},
  \bibinfo {author} {\bibfnamefont {C.~A.}\ \bibnamefont {Arg\"uelles}},
  \bibinfo {author} {\bibfnamefont {M.}~\bibnamefont {Bustamante}}, \ and\
  \bibinfo {author} {\bibfnamefont {A.~C.}\ \bibnamefont {Vincent}},\ }\href
  {\doibase 10.1088/1475-7516/2021/04/054} {\bibfield  {journal} {\bibinfo
  {journal} {JCAP}\ }\textbf {\bibinfo {volume} {04}},\ \bibinfo {pages} {054}
  (\bibinfo {year} {2021})},\ \Eprint {http://arxiv.org/abs/2012.12893}
  {arXiv:2012.12893 [hep-ph]} \BibitemShut {NoStop}%
\bibitem [{\citenamefont {Anchordoqui}\ \emph
  {et~al.}(2014{\natexlab{a}})\citenamefont {Anchordoqui}, \citenamefont
  {Goldberg}, \citenamefont {Lynch}, \citenamefont {Olinto}, \citenamefont
  {Paul},\ and\ \citenamefont {Weiler}}]{Anchordoqui:2013qsi}%
  \BibitemOpen
  \bibfield  {author} {\bibinfo {author} {\bibfnamefont {L.~A.}\ \bibnamefont
  {Anchordoqui}}, \bibinfo {author} {\bibfnamefont {H.}~\bibnamefont
  {Goldberg}}, \bibinfo {author} {\bibfnamefont {M.~H.}\ \bibnamefont {Lynch}},
  \bibinfo {author} {\bibfnamefont {A.~V.}\ \bibnamefont {Olinto}}, \bibinfo
  {author} {\bibfnamefont {T.~C.}\ \bibnamefont {Paul}}, \ and\ \bibinfo
  {author} {\bibfnamefont {T.~J.}\ \bibnamefont {Weiler}},\ }\href {\doibase
  10.1103/PhysRevD.89.083003} {\bibfield  {journal} {\bibinfo  {journal} {Phys.
  Rev. D}\ }\textbf {\bibinfo {volume} {89}},\ \bibinfo {pages} {083003}
  (\bibinfo {year} {2014}{\natexlab{a}})},\ \Eprint
  {http://arxiv.org/abs/1306.5021} {arXiv:1306.5021 [astro-ph.HE]} \BibitemShut
  {NoStop}%
\bibitem [{\citenamefont {Anchordoqui}\ \emph
  {et~al.}(2014{\natexlab{b}})\citenamefont {Anchordoqui} \emph
  {et~al.}}]{Anchordoqui:2013dnh}%
  \BibitemOpen
  \bibfield  {author} {\bibinfo {author} {\bibfnamefont {L.~A.}\ \bibnamefont
  {Anchordoqui}} \emph {et~al.},\ }\href {\doibase 10.1016/j.jheap.2014.01.001}
  {\bibfield  {journal} {\bibinfo  {journal} {JHEAp}\ }\textbf {\bibinfo
  {volume} {1-2}},\ \bibinfo {pages} {1} (\bibinfo {year}
  {2014}{\natexlab{b}})},\ \Eprint {http://arxiv.org/abs/1312.6587}
  {arXiv:1312.6587 [astro-ph.HE]} \BibitemShut {NoStop}%
\end{thebibliography}%

\end{document}